\documentclass[useAMS,usenatbib,usegraphicx]{mn2e} 
\usepackage{epsfig} 
 
\DeclareRobustCommand{\ion}[2]{%
\relax\ifmmode 
\ifx\testbx\f@series 
{\mathbf{#1\,\mathsc{#2}}}\else 
{\mathrm{#1\,\mathsc{#2}}}\fi 
\else\textup{#1\,{\mdseries\textsc{#2}}}%
\fi}

\newcommand{\vsini}{$v \sin i$} 
\usepackage[usenames]{color} 
 
\newcommand{\changeb}{\textcolor{Black} }

\def\kms{\hbox{$\,{\rm km}\,{\rm s}^{-1}$}} 
\def\ms{\hbox{$\,{\rm m}\,{\rm s}^{-1}$}} 
\def\ha{\hbox{$\,{\rm H}\alpha$}} 
\def\cd{\hbox{$\;{\rm c}\,{\rm d}^{-1}$}}

\def\degr{\hbox{$^\circ$}} 
 
\def\halpha{\hbox{${\rm H}\alpha$}}

\def\kms{\hbox{$\,{\rm km}\,{\rm s}^{-1}$}}

\title[3-D study of the photosphere of HD\,99563: I. Pulsation analysis]
{A 3-D study of the photosphere of HD\,99563: I. Pulsation analysis\thanks{Based
on observations collected at the European Southern Observatory, Paranal, Chile
(programmes 072.D-0138, 078.D-0192), and at the Subaru Telescope, which is
operated by the National Astronomical Observatory of Japan (programme
S07A-005).}}
\author[L. M. Freyhammer et al.]
 {L. M. Freyhammer$^{1}$\thanks{E-mail: lmfreyhammer\,@\,uclan.ac.uk},
 D. W. Kurtz$^{1}$,
 V. G. Elkin$^{1}$,
 G. Mathys$^{2}$,
 \newauthor{I. Savanov$^{1}$, W. Zima$^{3}$, H. Shibahashi$^{4}$ and K. 
Sekiguchi$^{5}$}\\
 $^{1}$Centre for Astrophysics, University of Central Lancashire, Preston
 PR1 2HE\\
 $^{2}$European Southern Observatory, Casilla 19001, Santiago 19, Chile\\
 $^{3}$Instituut voor Sterrenkunde, K.U. Leuven, Celestijnenlaan 200D, 3001
 Leuven, Belgium\\
 $^{4}$Department of Astronomy, {University of Tokyo}, Tokyo 113-0033, Japan\\
 $^{5}$National Astronomical Observatory of Japan, Mitaka, Tokyo 181-8588, Japan\\
 }

\begin{document}

\date{Draft \today ; Accepted . Received ; in original form }

\pagerange{\pageref{firstpage}--\pageref{lastpage}} \pubyear{2008}

\maketitle

\label{firstpage}

\begin{abstract}
We have used high-speed spectroscopy of the rapidly oscillating Ap (roAp) star 
HD\,99563 to study the pulsation amplitude and phase behaviour of elements in its 
stratified atmosphere over one 2.91-d rotation cycle. We identify spectral 
features related to patches in the surface distribution of chemical elements and 
study the pulsation amplitudes and phases as the patches move across the stellar 
disk. The variations are consistent with a distorted nonradial 
dipole pulsation mode. We 
measure a 1.6\,\kms\ rotational variation in the mean radial velocities of 
H$\alpha$ and argue that this is the first observation of H$\alpha$ abundance 
spots caused by He settling through suppression of convection by the magnetic 
field on an oblique rotator, in support of a prime theory for the excitation 
mechanism of roAp star pulsation. We demonstrate  
that HD\,99563 is the second roAp star 
to show aspect dependence of blue-to-red running wave line 
profile variations in Nd\,\textsc{iii} spots. 

\end{abstract}

\begin{keywords}
stars: individual: HD\,99563
-- stars: pulsating
-- stars: magnetic fields
-- stars: atmospheres
-- stars: chemically peculiar
\end{keywords}

\section{Introduction}
\subsection{The Ap stars}

Ap stars are chemically peculiar stars that range from early-B to early-F spectral 
types with the majority having detectable magnetic fields with strengths of a few 
$10^2$ to a few $10^4$\,G (\citealt{bychkovetal03}, according to whom $\sim55$ per 
cent have mean longitudinal fields greater than 400\,G). The study of Ap stars 
with strong magnetic fields has ramifications for several other branches of 
astrophysics. The interaction among strong magnetic fields, atomic diffusion and 
energy transfer in (or above) the upper atmospheres of non-degenerate stars has 
direct implications for observed stellar abundances and for the instability of 
stellar pulsations in $\beta$\,Cephei and sdB stars. The very different magnetic 
field strengths, orientations and geometries, as well as the great variety of 
surface abundance distributions for Ap stars, therefore makes them particularly 
informative. The magnetic field geometry, determined from maps of longitudinal 
magnetic field strength versus rotation phase, generally agrees with an 
axisymmetric dipole, although strong deviations from this do occur, e.g. the 
quadrupole field configuration of HD\,37776 (\citealt{thompson85}). Typically, 
the magnetic axis is significantly inclined to the rotation axis, hence these 
stars are called oblique rotators.

Many Ap stars are known to have surface spots, indirectly observable from 
photometric light variations or spectroscopic line strength variations over a 
rotation cycle. These stars are known as $\alpha^2$\,CVn stars after the prototype 
of that name \citep{pyper69}. Their light and line strength variations are a 
consequence of abundance spots that are often 
associated with the magnetic poles. The rotation 
period of Ap stars is typically relatively long -- of the order of several days to 
years, and even in some cases decades, as determined from their 
magnetic field variations. 
Slow rotation is a requirement for the atmospheric atomic diffusion that 
gives rise to the observed spectral peculiarities, since rapid rotation generates 
turbulent meridional circulation that inhibits diffusion. Interestingly, atomic 
diffusion processes can give rise to observably stratified atmospheres in some Ap 
stars, and probably all rapidly oscillating Ap (roAp) stars. Examples of studies 
of atmospheric abundance stratification in the cool roAp star subgroup are given 
by \cite{wadeetal01,ryabchikovaetal02,ryabchikovaetal05}. 

These studies are in agreement with each other. In general, 
Fe is concentrated by 
gravitational settling in the deeper observable layers in the photosphere around 
$\log \tau_{5000} \sim -0.5$; the Pr and Nd line-forming layers are concentrated 
by radiative levitation high in the atmosphere at $\log \tau_{5000} \le -4$, a 
level that is chromospheric in the Sun, and in the roAp stars is above the 
line-forming layer of the narrow core of the H$\alpha$ line at $-4 \le \log 
\tau_{5000} 
\le -2$, which is itself above the Fe line-forming layer. 
Many other elements show similar behaviour to these examples. 

Doppler imaging studies of Ap stars show that the spots associated with the 
magnetic poles are typically caused by abundance concentrations of rare earth 
elements. Overabundances globally are several orders of magnitude compared to 
normal abundance stars (e.g., \citealt{adelman73}), and may reach a factor of 
$10^8$ in the spots, as shown by the Doppler imaging study of the roAp star 
HR\,3831 by \cite{kochukhov04}, and as shown in this paper for HD\,99563. 
As a result of these 
strong abundance anomalies, 
line-blocking changes the temperature gradient at the magnetic poles 
of Ap stars; flux is redistributed usually from the blue to the red, so that 
rotational light variations in visible light 
occur in antiphase for short and long wavelength 
observations, along with obvious spectral line strength variations. Importantly, 
the rotational variations 
-- whether magnetic, spectroscopic or photometric -- 
allow the precise determination of the stellar rotation period. 

With the surface of some Ap stars resolvable by Doppler imaging, and atmospheric 
depth being resolved because of stratification, Ap stars -- and particularly the 
roAp stars -- offer the only opportunity for 3-D studies of stellar atmospheres 
for any stars other than the Sun.

\subsection{The roAp stars}

A subgroup of Ap stars, the rapidly oscillating Ap stars (roAp), exhibit rapid 
oscillations with periods of $5-21$\,min. Their frequencies are considerably 
higher than those typical for $\delta$\,Sct stars with whom they overlap in 
location inside the classical instability strip of the HR diagram. At present, 
only 40 roAp stars are known (see, e.g., \citealt{kurtzetal06b,gonzalezetal08}), 
although several photometric surveys have searched for rapid pulsation in Ap 
stars, such as \citet{nelsonetal93, martinezetal94, handleretal99, ashokaetal00, 
weissetal00, dorokhovaetal05}. For a (non-exhaustive) list of spectroscopic 
studies of roAp stars, see Kurtz, Elkin \& Mathys (2006a). 

The roAp class characteristics are therefore poorly constrained. As they share the 
same region of the HR diagram with the (photometrically established) 
non-oscillating Ap (noAp) stars, 
the selection of new roAp candidates is difficult and 
mostly leads to null-results for both spectroscopic studies (see, e.g., 
\citealt{freyhammeretal08a}; Freyhammer, Elkin \& Kurtz 2008b; 
\citealt{elkinetal08a}) and for the 
photometric surveys listed above. The pulsations of roAp stars are described well 
by the oblique pulsator model, where the pulsation axis is assumed to be aligned 
with an oblique magnetic field axis (\citealt{kurtz82}, \citealt{saio05}), or 
derived to be offset from both the rotational and magnetic axes \citep{bigot02}. 

\citet{balmforthetal01} used this model and assumed that convection was suppressed 
locally in the magnetic polar regions. They showed that pulsation aligned with the 
magnetic poles is unstable to the axisymmetric, nonradial high-order 
modes observed in 
roAp stars, mainly through excitation by the opacity ($\kappa$) mechanism 
acting in the hydrogen ionization zones in these regions. They also considered the 
effect of local He settling, and concluded it is not the primary process 
responsible for the high-frequency pulsation. Nevertheless, their model finds He 
to be deficient in the observable atmosphere at the magnetic poles. While He is 
not directly observable at visible wavelengths 
at the effective temperatures of the roAp stars ($6600 \le 
T_{\rm eff} \le 8200$\,K) its gravitational settling implies that the magnetic 
poles are hydrogen-rich. We return to this point in Section~\ref{hcore} where we 
show evidence to support it. 

Statistical studies of the roAp stars can be intriguing and informative, e.g. in 
the comparison of the roAp and noAp stars by \citet{hubrig00}. But with only 40 
such stars known the class characteristics are still not well-defined. Therefore, 
the greatest progress at present is to be made by in-depth studies of individual 
roAp stars, such as that for HR\,3831 (e.g. \citealt{kurtzetal97};
\citealt{kochukhov06}) and HR\,1217 (\citealt{kurtzetal05a};
\citealt{ryabchikovaetal07}). 

This is reminiscent of the related problem in the 
pulsational studies of $\delta$\,Sct stars. Despite being 
the most abundant type of variable star, the major breakthroughs have been 
obtained through individual studies such as that 
of FG\,Vir \citep{bregeretal05}. \changeb{For the $\delta$\,Sct stars, as for the roAp stars,  
the question of when pulsation modes }
are exited or not remains unanswered for physically similar stars -- why do only 
some of these stars pulsate and what is the underlying physics? The many null 
results among roAp candidate stars strongly suggest that either the pulsation 
properties are not fully known (i.e. amplitude ranges or pulsation geometry) or 
the selection of candidates is too coarse. Individual studies of known roAp stars 
are therefore important (presently the only option at hand) to study the roAp 
properties in detail. Some examples of enlightening studies include that of 
$\gamma$\,Equ that clearly demonstrated the trends for 
pulsational amplitude with vertical stratification of 
chemical elements (\citealt{savanov99};  \citealt{kochukhovetal01}), Doppler 
imaging of HR\,3831 (Kochukhov 2004, 2006), and the study of line bisector 
variability by \cite{kurtzetal06c} and a new type of upper atmospheric pulsation 
(\citealt{kurtzetal06a}; \citealt{kurtzetal07}).

\subsection{The roAp pulsations}

To first order, pulsations in many roAp stars are oblique axisymmetric dipole 
($l=1$, $m=0$) p~modes. Nevertheless, it has been clear for decades that normal 
modes described by single spherical harmonics are insufficient to explain fully 
the mode geometry in these stars. Uniquely, the oblique modes of the roAp stars 
allow them to be viewed from varying aspect, giving detailed information about the 
mode geometry. An early example was the study of HR\,1217 by \cite{kurtzetal89} 
who showed the presence of alternating even and odd degree 
p~modes, with the odd modes 
being consistent with oblique dipole modes, and the even modes not being 
consistent with either radial or quadrupole modes. This was confirmed by the more 
extensive Whole Earth Telescope study of this star by \cite{kurtzetal05a}. Even 
for roAp stars with single pulsation modes, in the best observed cases -- such as 
HR\,3831 -- it can be \changeb{shown that} the single mode is a distorted oblique 
dipole mode (\citealt{kurtzetal97}; \citealt{kochukhov06}). 

Any deviation from spherical symmetry for a pulsating star will give rise to 
distorted modes. For the roAp stars the dominant effect is the magnetic field, 
with contributions from the non-uniform abundance distributions and stellar 
rotation. The non-uniform abundance distributions are particularly important for 
spectroscopic studies -- such as the one in this paper on HD\,99563, and that of 
\cite{kochukhov06} for HR\,3831 -- where concentrations of ions in spots mean that 
the pulsation mode is non-uniformly sampled with the rotation. However, 
photometric studies that more uniformly sample the mode geometry with rotation 
also show the distortions of the modes. 

The effects of the magnetic field on the oscillations of roAp stars have been 
extensively studied theoretically, and there are ongoing investigations into this 
complex problem (\citealt{dzgoode96}; 
\citealt{bigotetal00}; \citealt{cunhagough00}; 
\citealt{bigot02}; \citealt{saioetal04}; \citealt{saio05}; \citealt{cunha06}; 
\citealt{sousacunha08}). While generally the magnetic field effect on the 
oscillations is expected to be small, the theoretical results do lead to 
expectations of distorted modes. For example, \citet{saioetal04} studied the 
influence of the magnetic field on the pulsation geometry and found that not only 
can a dipole magnetic field explain the suppression of the observationally absent 
$\delta$\,Sct modes, but it also leads to insufficiency of a single spherical 
harmonic to describe the pulsations angular dependency; a series of spherical 
harmonics of different degrees $l$ are required. Inclusion of magnetic fields in 
the study and interpretation of roAp pulsations is therefore necessary and 
constitutes `a formidable mathematical problem' \citep{saioetal04}.

\citet{saio05} continued the study of the effects of a dipole magnetic field on 
roAp pulsations and showed again that the field stabilised low-order p~modes (such 
as those excited in $\delta$\,Sct stars) for fields stronger than 1\,kG, but 
that nonradial, high order distorted dipole, or quadrupole, p~modes remain 
overstable and are most likely to be excited in roAp stars. This is in excellent 
agreement with the current observations, as well as with the mounting 
observational and theoretical evidence for observable pulsation nodes in and above 
the photosphere in the magneto-acoustic layer. Further, and most 
interestingly, Saio modelled the latitudinal amplitude dependence for the roAp 
case of HR\,3831 and found reasonable agreement with the observations by 
\cite{kochukhov04}. 

Saio's non-adiabatic analysis predicts (his figure 10) a latitudinal dependence of 
amplitude that does not differ from a purely dipole mode photometrically, whereas 
spectroscopy's ability to distinguish potentially between horizontal and radial 
pulsation components may detect the relatively smaller horizontal component when 
studying stellar spots close to (but not necessarily at) the magnetic poles. 
Saio's figure 10 demonstrates a reasonable agreement with observed latitudinal 
velocity amplitude variation for HR\,3831, but predicts higher amplitude toward 
the magnetic axis than is observed. However, the formation layer of the Nd lines 
used by Kochukhov in his observations is far above the boundary of Saio's models. 
\citet{kochukhov06} also did not take into account the inhomogeneous Nd surface 
distribution.

The oblique pulsator model describes how nonradial pulsation modes seen from 
varying aspect with the stellar rotation show amplitude and phase modulation (see 
the overview by \citealt{kurtzetal00}) that results in a frequency multiplet in an 
amplitude spectrum, centred on the pulsation frequency $\nu$ and separated by {\it 
exactly} the rotation frequency $\nu_{\rm rot}$. This exact splitting by the 
rotation frequency allowed Kurtz (1982; see also \citealt{kurtzetal00}) to  
rule out the possibility that the observed multiplets can be explained by 
rotationally perturbed $m$-modes ($m$ being the azimuthal order of a mode), which 
is the typical interpretation of frequency splitting observed in other 
(non-oblique) nonradially oscillating stars, as such rotational 
multiplets are split by 
$(1-C_{nl})\Omega$, where $\Omega$ is the rotation frequency,  
$C_{nl}$ is a small constant dependent on the stellar structure, 
$n$ is the overtone of the mode and $l$ is its degree. 

For oblique pulsation the amplitude ratio of the multiplet components contains 
information about the pulsation geometry (assumed to be aligned with the magnetic 
geometry). In the case of a simple oblique dipole mode 
$(A_{+1}+A_{-1})/A_0=\tan{i}\tan{\beta}$, where $i$ is the 
inclination angle of the rotation 
axis, $\beta$ is the obliquity of the pulsation/magnetic axis relative to the 
rotation axis, and $A_0, A_{+1}$, $A_{-1}$ are the observed amplitudes of the 
frequency triplet. For modes that are not purely dipolar, or vary in structure 
with atmospheric depth in the stratified roAp stars atmospheres, this constraint 
can depend on the atmospheric depth of formation of the spectral lines studied 
(see, e.g., \citealt{kochukhov06}).

Despite the complex nature of roAp pulsations in the presence of magnetic fields 
and an oblique geometry, the observational data have rich potential for 
asteroseismic inference. Photometrically, multisite \changeb{observations and space-borne 
observations} allow the resolution of both the large and small frequency spacings 
and through them constrain the stellar dimensions and the interaction of the 
magnetic field and rotation with the pulsation modes. High-resolution time series 
spectroscopy allows the study of the pulsation geometry over the stellar surface 
(through line profile variability) and also vertically in the stellar atmosphere, 
through radial velocity measurements at different line bisectors, or through lines 
of elements located at different depths in the stratified atmosphere (e.g., 
\citealt{kurtzetal06c,ryabchikovaetal07}). It is clear that the observations 
have begun to resolve the magneto-acoustic nature of the pulsations in these stars 
high into the atmosphere to levels that are unobservable in any other star but the 
Sun. Radial nodes are resolved in some stars such as HD\,137949 
(Mkrtichian, Hatzes \& Kanaan 2003; Kurtz, Elkin \& Mathys 2005c) 
and possibly HD\,99563 
\citep{elkinetal05}. Observed variation of pulsation phase with atmospheric depth 
shows the presence of running waves, and there is even an indication that shock 
waves are observed in the upper atmosphere of $\gamma$\,Equ  
(\citealt{shibahashietal08}), 
and by inference other roAp stars with similar line profile variations.  

\subsection{HD\,99563}

HD\,99563 was discovered to be a roAp star by \cite{dorokhovaetal98}, pulsating 
in a 10.70-min mode. It is a visual binary with a 1.2\,mag fainter secondary at 
1.79\,arcsec separation \citep{fabriciusetal00}. A spectroscopic study by 
Elkin et al. (2005) confirmed the monoperiodic pulsation found 
from photometry, and 
found the pulsations to have extremely high radial velocity amplitudes for a 
roAp star, such as 
2.6\,\kms\ for \halpha\ and up to 5\,\kms\ for Eu and Tm. For sake of 
clarity, we \changeb{emphasize} that {\em monoperiodic} refers to a single (physical) 
pulsation mode -- e.g., the frequency quintuplet detected by \citet{handleretal06} 
(actually 5 components of a frequency septuplet) 
are all associated with only a single mode 
with pulsation frequency $\nu$; the 
rest of the quintuplet describes the amplitude and phase modulation of the mode. 

Elkin et al. (2005) measured \changeb{a projected rotation velocity of} \vsini$~=28.5\pm1.1$\,\kms\ from 
a high-resolution spectrum, while from narrow-band Str\"omgren 
photometric indices they estimated $T_{\rm  eff} = 7700$\,K and $\log g = 4.2$ 
(cgs). Furthermore, they studied variations in 
amplitude and phase for line bisectors, and from element to element. The 
demonstrably high amplitudes (and thus high signal-to-noise, $S/N$) and rich and 
varied information content from using lines of different elements or with line 
depth, show HD\,99563 to be a particular interesting roAp star for detailed study.

\citet{handleretal06} carried out a photometric multisite campaign and found an 
equally spaced frequency quintuplet in HD\,99563, for the pulsation frequency 
$\nu=1.5576539\pm0.0000007$\,mHz. They determined the rotation period to be 
$P_{\rm rot} = 2.91179\pm0.00007$\,d and a time of pulsation amplitude maximum to 
be HJD\,245\,2031.29627. The photometry showed, through application of the 
axisymmetric spherical harmonic decomposition method of \citet{kurtz92}, that the 
mode of HD\,99563 is dominated by a dipole component, with some contribution of an 
$l=3$ component (i.e., the mode is a distorted dipole). 

Handler and collaborators combined photometry, a Hipparcos trigonometric parallax 
and the temperature and surface gravity estimate by Elkin et al. (2005) to 
constrain the basic parameters of HD\,99563. Under the assumption 
that the secondary star 
in the visual binary is a physical binary component and that the wide orbit 
indicates the stars have evolved separately, the estimated parameters of both 
stars must fit two stellar evolutionary models of the same age. By adopting 
$T_{\rm  eff} =7900 \pm 300$\,K and $\log g = 4.152$, they found for the primary 
$\log L/{\rm L}_\odot=2.03$, $R/{\rm R}_\odot=2.38$, an inclination angle of the 
stellar rotation axis $i=43\fdg6 \pm 2\fdg1$, and a magnetic obliquity 
$\beta=86\fdg4 \pm 0\fdg3$. This makes HD\,99563 especially interesting because of 
its high pulsation amplitude, its rotation that is fast enough for Doppler 
imaging, and its (for a roAp star) rare favourable orientation where both magnetic 
poles become visible throughout a rotation cycle.

\citet{hubrigetal04} were the first to measure the magnetic field of HD\,99563, 
finding a longitudinal field of $-688\pm145$\,G on HJD\,245\,2494.483, or rotation 
phase 0.08 with respect to Handler et al.'s pulsation ephemeris. 
\citet{hubrigetal06} published another two measurements at rotation phases 0.07 
and 0.09 showing $-235\pm73$ and $670\pm37$\,G, respectively. Further, using the 
Russian 6-m telescope at the Special Astrophysical Observatory of the Russian 
Academy of Sciences (SAORAS), \citet{elkinetal08b} obtained 6 additional magnetic 
longitudinal field measurements near rotation phases 0.25, 0.5 and 0.75. With all 
9 available measurements, these authors used the well-known relation
for a dipolar magnetic field:
\begin{equation}
\label{eq:mag1}
B_{\rm l} \propto B_{\rm p} \cos \alpha
\end{equation}
\noindent where
\begin{equation}
\label{eq:mag2}
\cos \alpha = \cos i \cos \beta + \sin i \sin \beta \cos \Omega t,
\end{equation}
\noindent $\alpha$ is the angle between the magnetic pole and the
line-of-sight, 
$i$ is the rotational inclination, $\beta$ is the angle between the
rotation axis and the magnetic axis, $\Omega$ is the rotation frequency,
$B_{\rm l}$ is the longitudinal magnetic field strength and $B_{\rm p}$ 
is the polar field strength.

As can be seen from Eq.\,\ref{eq:mag2}, the mean magnetic field strength 
$\langle B_{\rm l} \rangle \propto \cos i \cos \beta$
and the amplitude of the magnetic field variations
$ A_{B_{\rm l}} \propto \sin i \sin \beta$. 
Thus $\tan i \tan \beta = A_{B_{\rm l}}/\langle B_{\rm l} \rangle$,
from which $\beta$ can be constrained when $i$ is known. Using
the measured values for HD\,99563, 
$\langle B_{\rm l} \rangle = 21 \pm 58$\,G and $A_{B_{\rm l}} = 
701\pm114$\,G, \citet{elkinetal08b} find an
inclination of $i = 43\fdg5$, and a magnetic obliquity $\beta = 88\degr$ 
(84--90\degr\ for a 1$\sigma$ range), which are 
similar to those determined by \citet{handleretal06}. The magnetic minimum 
coincides with the moment of maximum modulated pulsation amplitude ($\varphi_{\rm 
rot}=0$ hereafter), and similarly for the maximum half a rotation phase later 
($\varphi_{\rm rot}=0.5$). This is consistent with both magnetic poles 
being viewed from a similar angle and coming into view 
for almost the same duration 
(requiring that either $i$ or $\beta$ be close to $90^\circ$), 
and with the pulsation axis and magnetic axis being aligned. 

\citet{kochukhov04} used Doppler imaging to construct the first tomographic maps 
of 
pulsation velocity, surface abundance and magnetic field map for the roAp star 
HR\,3831. With observations obtained over a wide range in rotation phase for 
HD\,99563, it will be possible to do this with higher precision for this star 
because of its very high amplitudes, and it will be possible to look 
simultaneously at the additional dimension of the atmospheric depth dependence of 
the velocities. Kochukhov, as a first approximation, treated the analysis of 
HR\,3831 as if the atmosphere has a single pulsation layer. Observations of 
HD\,99563 and other roAp stars show the situation to be much more complex than 
this. Disentangling the vertical and horizontal pulsation characteristics is a 
challenge, but one with rich rewards for the fields of asteroseismology, pulsation 
theory, magnetohydrodynamics, stellar atmosphere theory and atomic diffusion.

HD\,99563 is one of only a few known roAp cases where tomographic mapping through 
Doppler imaging is possible, based on rotation rate and the orientation of the 
magnetic field. Therefore, observations over three nights 
(most of the 2.91-d rotation period) were scheduled 
from two sites simultaneously -- 
Subaru on Mauna Kea and VLT on Cerro Paranal -- 
to attempt to collect spectra at 
high resolution in both time and dispersion to obtain a 
first picture of this star's photosphere, and to determine precisely its physical 
parameters. Theoretically, the scheduled observations  
would have given us 70 per cent coverage of the rotation cycle, hence of the 
visible stellar surface. 
But due to bad weather, only one site contributed significantly to the 
pulsation study, and only some 40 per cent rotational coverage was attained. 

In this first of a series of papers on HD\,99563, 
we are studying the horizontal 
pulsation geometry with the stellar rotation. A Doppler imaging 
study of the surface abundance distribution of many ions 
and detailed modelling of the 
horizontal and vertical pulsation geometry \changeb{is} in preparation.

\section{Observations and data reduction}

\begin{table} 
 \begin{minipage}{81mm} 
\caption{{\normalsize \label{tab:obslog}Journal of VLT observations indicating 
nightly observing date (UT), rotation phase with reference to a time of 
spectroscopic pulsation maximum, HJD\,245\,4171.43328,  
Heliocentric Julian Date (HJD) range, typical exposure time, number of collected 
spectra, and mean $S/N$. The $S/N$ \changeb{ratios are based on the random noise 
in the continuum of consecutive spectra}, or for single spectra 
based on predicted ratios. The 2004 data were re-reduced in same manner as the new 
data, and included in the analysis as they supplement the covered rotation phases. 
An additional 35 spectra were obtained with the HDS spectrograph at the Subaru 
telescope on 2007 March 12 and March 13 and combined into two average spectra at 
rotation phases $\varphi_{\rm rot}=0.23$ and 0.56. }} 
\begin{tabular}{@{}l@{~~}l@{~~}c@{~}c@{\,}c@{}r@{}c@{}} 
\hline 
Date (UT) & Rotation & {HJD range} &$t_{\rm exp}$& $n$ &\multicolumn{2}{c}{$S/N$} \\ 
 & phase $\varphi_{\rm rot}$ & ({\footnotesize $-$245\,0000}) &(s) & 
& {\footnotesize $\lambda${\scriptsize 5100}} & \,{\footnotesize $\lambda${\scriptsize 6400}} \\ \hline 
2004 Mar 06 & 0.937--0.966 & 3070.60--3070.69 &40 & 111 & 120 & 95 \\ 
2007 Mar 12 & 0.021--0.150 & 4171.53--4171.91 &40 & 518 & 75 & 55 \\ 
2007 Mar 13 & 0.365--0.494 & 4172.52--4172.90 &40/80 & 417 & 85 & 60 \\ 
2007 Mar 14 & 0.708--0.837 & 4173.52--4173.90 &65 & 372 & 110 & 75 \\ 
 \hline 
\hline 
\end{tabular} 
\end{minipage} 
\normalsize 
\end{table} 
 
Time-series spectroscopy observations were collected on Cerro Paranal with the VLT 
UV-Visual Echelle Spectrograph (UVES) during three nights in 2007 March; a journal 
of observations is given in Table\,\ref{tab:obslog}. UVES is an echelle 
spectrograph and we used the settings for the wavelength range $4970-7010$\,\AA\ 
(with a 60-\AA\ wide gap near 6000\,\AA) at $R=110\,000$. The UVES instrument was 
used with an image slicer and a 0.3\,arcsec slit. Typical exposure times were 40 
and 65\,s, and 80\,s at high airmass on the second night. With $\sim26$\,s readout 
and overhead time, a time resolution of $66-106$\,s was obtained.

Table\,\ref{tab:obslog} lists the spectra collected each night. With the 2.91-d 
rotation period of HD\,99563, the 3 night run was scheduled to cover most of a 
rotation cycle of HD\,99563 (rotational phases $0.02-0.84$) 
and was coordinated with a run 
using the High Dispersion Spectrograph (HDS) on the 8.2-m Subaru telescope at 
Mauna Kea to obtain the rotational phases coinciding with daytime on Paranal. 
However, humid weather conditions at Mauna Kea resulted in only 35 sporadic 
observations that are not included in the frequency analysis in 
Sect.\,\ref{sec:linana}, but were instead combined into two average spectra at 
rotation phases 0.23 and 0.56 and used in Doppler imaging to produce surface maps 
for chemical elements, only two of which are shown in this paper (see 
Fig.\,\ref{fig:ndDI} and Fig.\,\ref{fig:euDI} below); a full Doppler imaging 
analysis will be presented in a future publication. 
For details on the Doppler imaging programme and production of the 
tomographic maps see \citet{savanovetal05}. 

The spectra were reduced to 1-D with the pipelines provided by ESO with the 
standard calibration data (bias, flatfield and Thorium-Argon wavelength reference 
spectra). The $S/N$ was 27 per cent lower than predicted with ESO's online 
exposure time estimation tool. The observing conditions were in general good, all 
3 nights were clear and dry, with typical seeing of $0.5-0.8$\,arcsec, 
except for the first night where seeing was $0.7 - 1.1$\,arcsec 
during the first half of the night and 
then deteriorated to $0.8 - 2.4$\,arcsec thereafter.

Special effort was therefore made to optimise the reduction. We used the {\sc 
UVES pipeline} 2.9.7 (based on {\sc MIDAS}) and applied minor additional 
\changeb{corrections:} We smoothed out a CCD bias pattern for every 3.5 columns (RED 
upper CCD), inserted 2 columns missed during CCD readout (with minor effects on 
the 2-D wavelength calibration), and fixed the rebinning step sizes of the 1-D 
spectra to 0.01787\,\AA\ (RED upper CCD) and 0.01526\,\AA\ (RED lower CCD). To 
eliminate wavelength drifts, wavelength calibrations were made to the nearest 
Th-Ar reference spectrum (obtained \changeb{every 3\,hours).}

Continuum rectification of the 1-D spectra was performed iteratively: individual 
spectra were normalised to a run master from the most stable night (starting on 
2007 March 12) using a 4th-order polynomial, then these were normalised to master 
spectra of individual nights. After inspection, subsets of the resulting spectra 
were corrected for remaining slopes. The nightly masters were rectified 
through iterative normalisation using spline fitting to pseudo-continuum windows 
identified from comparison to a model spectrum. Features due to cosmic rays and 
blemishes were removed with a semi-automatic cutting procedure. Barycentric 
corrections, which changed by about 1.0\,\kms\ over a single night, 
were registered 
and applied in the period analysis, but not in the wavelength calibration and 
final rebinned 1-D spectra.

Table\,\ref{tab:obslog} gives the estimated signal-to-noise ratio ($S/N$)
 of the spectra 
based on random noise measured in difference spectra of pair-wise subsequent 
spectra. Because the 2007 data had 27 per cent lower $S/N$ than predicted, and 
also compared to the 2004 observations for the same exposure time, exposure times 
were increased to 60\,s most of the two last nights in 2007. This difference is 
not due to weather, as large seeing variations are found to have only small 
influence on the $S/N$ (as a consequence of the image slicer). 
Comparing the 2-D spectra of the two 
years confirms that fewer photons were collected 
in 2007 than in 2004. About 10 per 
cent variation in $S/N$ is due to the differences in airmass, since in 2007 we 
observed HD\,99563 during full nights. Part of the reason for the lower $S/N$ is 
from centring of the slicer on the object, but alone this cannot explain the loss. 
The spectra from the upper CCD, redward of 6000\,\AA, show a systematic noise 
pattern of undulations with $\sim3.4$ cycles per \AA, for stretches of 
$\sim$30\,\AA\ occurring every $\sim$67\,\AA, starting at 6070\,\AA. This 
coincides with the first half of each Echelle order. As HD\,99563 rotates 
relatively quickly, resulting in broad lines, we smoothed 1-D 
spectra of the upper CCD over $3-4$ pixels, 
which strongly suppressed the artefact pattern. 
Comparing our frequency analyses for lines from the lower and upper spectra gave 
no significant differences. We suspect the origin of this feature is a sort of 
amplification of the aforementioned 3.5 column pattern at the order merging 
regions. During the last night a 1\,h gap in the time series occurred when 
the image acquisition system failed to save the observations.

\section{Data analysis and results}

\subsection{Line analysis}
\label{sec:linana}

For the purpose of spectral line identification, a synthetic comparison spectrum 
was produced with {\sc SYNTH} \citep{piskunov92} using a Kurucz stellar 
atmosphere model. Atomic line data were taken from the Vienna Atomic Line Database 
(VALD, \citealt{kupkaetal99}) for ions with increased abundances, mainly for Nd, 
Pr, Sr, Cr and Eu. Other sources used for line data were the atomic database 
NIST (National Institute of Standards and 
Technology)\footnote{http://physics.nist.gov} 
and the Database on Rare Earth Elements at 
Mons University (DREAM\footnote{http://w3.umh.ac.be/$\sim$astro/dream.shtml}; 
\citealt{biemontetal99}) through its implementation in the VALD. The line 
selection and line analysis is fully based on lines identified in table\,1 of 
Elkin et al. (2005).

\subsubsection{Radial velocity shifts}
\label{sect-rvshift}

Because of the non-uniform surface distributions and 
stratification of ions in the atmospheres of Ap stars, 
high-overtone mode pulsation 
amplitude and \changeb{phase vary from} element to element, and even within individual 
lines as a function of line depth (atmospheric height). Line profiles are variable 
because of nonradial pulsation, non-uniform abundance of elements over the surface 
because of spots, and non-uniform abundance with depth because of stratification. 
It is our intention with this and future studies of the data presented here to 
disentangle these contributions to extract 3-D radial velocity maps of the stellar 
pulsation geometry. 

{\sc IDL} tools for line measurements and analyses were used as described in 
\citet{freyhammeretal08a}. Precise radial velocity shifts were measured for 
many spectral lines with the centre-of-gravity (CoG) method and by fitting with 
Gaussian profiles. We made our measurements with respect to 
the continuum -- normalised to unity -- in the selected regions of the 
measured lines, which gave smaller amplitudes for the CoG procedure. 
This method, 
however, was preferred for pulsation studies because of its better ability to deal 
with blended line profiles. For strong and isolated lines, the two methods were 
comparable. However, in some cases Gaussian fitting was used in automated 
measurements as it is more stable when the whole line profile shifts considerably 
due to, e.g., rotational mean radial velocity modulation (see Sect.\,3.2) when 
 one of the line wings goes outside the pre-defined wavelength window. 
The main difference of 
the two methods is their estimates of pulsation amplitudes. The CoG method is very 
sensitive to continuum placement and more sensitive to subtle, intrinsic line 
profile variations of relatively weak, asymmetric lines.

The radial velocity measurements were automated with a wavelength window
defined for each line in a high $S/N$ average spectrum. That window was
shifted in wavelength to account for the barycentric correction for each
individual spectrum, then the profile was fitted with both a Gaussian and
by
CoG fitting; finally, barycentric corrections were added to the resulting
radial velocities. The two radial velocity series were then compared.

Because of the occurrence of substructure in absorption features of elements with 
inhomogeneous surface distributions, we used two approaches to study line 
variability: 1) to study an element's integrated radial velocity behaviour, we 
used the full line profile, whereas 2) for line profiles with 
sub-features caused by inhomogeneous (spotted) surface distributions, we 
measured each substructure in a line and followed it over the half a 
rotation cycle for which it was visible. 
The line windows for the `spot' features (two for Nd and 3 for Eu, 
see below in the text and Figs\,\ref{fig:ndpattern} and \ref{fig:eupattern} for 
clarification) were defined in 1\,hr sub-series of spectra (for a total of 32 time 
bins), rather than one window for all 1418 spectra. The former approach was useful 
for fitting multiplet terms (Sect.\,3.2), while the latter approach was 
particularly useful for studying pulsation amplitude and phase variations with 
rotation phase. 

\subsubsection{Frequency analyses}

Frequency analyses were performed using a Discrete Fourier Transform programme 
\citep{kurtz85} and the {\mbox{\sc PERIOD04}} \citep{lenzetal05} programme. The 
former fits cosines while the latter uses sines, so all phases determined with 
{\mbox{\sc PERIOD04}} were recalculated to match the cosine fitting in 
units of radians. The noise, $\sigma$, of fitted amplitudes and phases (see, e.g., 
Tables\,\ref{tab:frq} and \ref{tab:frq2}) was determined from least-squares 
fitting to the data following \citet{deeming75}; it is the standard 
deviation of one measurement with respect to the fitted function.

\begin{figure}
\includegraphics[height=0.45\textwidth, angle=90]{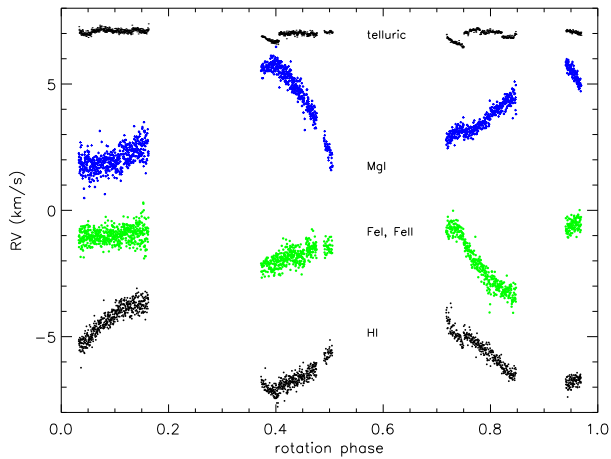}
\caption{\label{fig:rotvar}
Rotational radial velocity trends for different elements: telluric lines (2 lines 
combined), Mg (two Mg\,\textsc{i} lines combined), Fe (combined from 4 
Fe\,\textsc{i} and Fe\,\textsc{ii} lines), and the hydrogen core (prewhitened for 
pulsations). Barycentric corrections were applied to the radial velocities of 
stellar lines and are relative to those of the telluric lines. Radial velocity 
offsets were added to each series' ordinates for clarity. If the radial velocity 
variation with the rotation period 
\changeb{was} the result of reflex motion of HD\,99563 in a binary orbit with an unseen 
close companion, the amplitudes and trends of relatively homogeneously distributed 
elements, such as iron and hydrogen, should be the same; they are not. The 
radial velocities for the iron measurements that deviate most from those for 
H$\alpha$ at phase $0.94$ are from 2004, but there is also significant deviation 
at phases $0.02 - 0.17$ in the 2007 data. Mg shows 
significantly different trends from Fe and H, but this is most probably caused by 
a patchy surface distribution. Because of the non-uniform abundance distributions, 
orbital radial velocity shifts of HD\,99563 can only be ruled out for amplitudes 
significantly larger than those seen here. }
\end{figure}

Trends in the radial velocities with the rotation were strong for all lines, even 
\halpha\ and Fe (Fig.\,\ref{fig:rotvar}), indicating that HD\,99563 is very 
spotted. Linear and non-linear trends were fitted and removed with linear or 
polynomial least-squares fitting, and in a few cases also by prewhitening 
low frequency peaks from the 
radial velocity series. The origin of 
these trends are discussed further in Sect.\,3.2, but most of the contributions 
are stellar; with the 0.3-arcsec \changeb{slit, no image slicer} and seeing conditions of $0.9 - 
1.4$\,arcsec, the centring error for UVES is only $50-100$\,\ms\ 
\citep{bouchyetal2004}. Furthermore, a 1-mbar change in pressure may induce drifts 
of 90\,\ms. Therefore, instrumental shifts were less than $\sim$300\,\ms\ per 
night, depending on seeing and pressure. 

The telluric lines in Fig.\,\ref{fig:rotvar} show that the wavelength 
calibration sometimes has small offsets (up to 600 \ms) between reductions 
using different wavelength reference spectra. These are artefacts from 
not interpolating over all reference spectra for a night, an option not
available with the UVES pipeline, and then rebinning to same-width 
wavelength bins. However,
as the time scale of these offsets is very 
different from that of the pulsations (3\,hr as opposed to 11\,min), 
we did not correct this in the analysis. Telluric lines 
only provide accurate and precise velocity references within a 
few hundred \ms, as they are influenced by fast and changing wind speeds in 
the high layers of the Earth's atmosphere where they are formed.

\subsection{Hydrogen core}\label{hcore}

\citet{handleretal06} detected 10 frequencies in their time series multisite 
campaign $B$-band photometry of HD\,99563. These are comprised of 
 (Table\,\ref{tab:frq}) 
the pulsation frequency $\nu=1.557653\pm0.000007$\,mHz, the rotational split 
oblique dipole frequency quintuplet: $\nu\pm\nu_{\rm rot}$ and $\nu\pm3\nu_{\rm 
rot}$, and other harmonics or combination frequencies. \changeb{These authors identified
the rotation frequency as $\nu_{\rm rot}= 0.0039751$\,mHz and gave the
photometric ephemeris (in days) with respect to the time of photometric
modulated pulsation amplitude maximum as}: 
\begin{equation}
\label{eq:ephem0}
t({\rm max}) = {\rm HJD}\,245\,2031.29627 + 2.91179 \pm 0.00007\,E,
\end{equation}
\changeb{where $t($max$)$ gives the time in HJD corresponding to the given number of 
epochs $E$ elapsed
since the reference point in time. The decimal part of $E$ corresponds to the rotation
phase in the case of HD\,99563.   }
The pulsation frequency 
$\nu$ itself has the second lowest amplitude of the quintuplet. See Handler et 
al.'s figure 5 for a schematic view of their frequency quintuplet. 

The \ha-core line-forming region extends over a considerable vertical region of 
the chemically stratified photosphere, in contrast to other elements. Radial 
velocity measurements of this line therefore compare well to the photometry, and 
as \halpha\ for this star has a high pulsation amplitude in excess of 2\,\kms, we 
used this line for the initial frequency analysis. Because of the 
\changeb{limited} accuracy of the rotational period in  
Eq.\,\ref{eq:ephem0}, a reference epoch close to our 2007 observations was chosen 
at the moment of maximum of the modulated pulsation amplitude of \halpha:
\begin{equation}
\label{eq:ephem}
t({\rm max}) = {\rm HJD}\,245\,4171.43328\,+2.91179\,(\pm0.00007)\,E,
\end{equation}
while the value of the pulsation frequency $\nu$ by \citet{handleretal06} was 
revised within its error to $\nu=1.5576530$\,mHz in order to tie the new spectra 
together with the 2004 data set (see Sect.\,\ref{sec:halinana}). 

\subsubsection{Trends of the radial velocity shifts}
\label{sec:halinana}

\begin{figure}
\includegraphics[height=0.45\textwidth, angle=90]{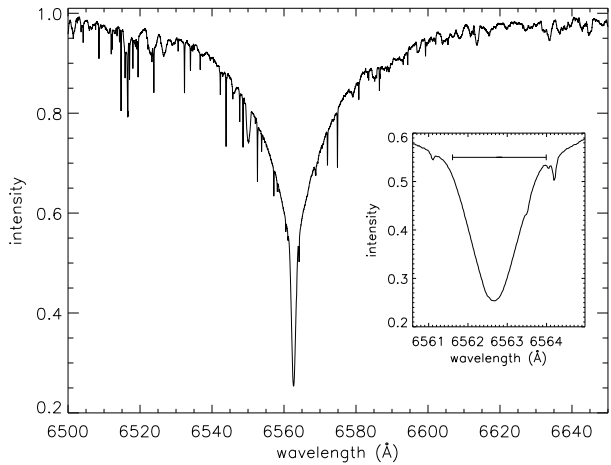}
\caption{\label{fig:haregion} Average spectrum of the \halpha\ region. The insert 
shows the line core, with the region measured for radial velocities 
marked with a bar. }
\end{figure}

\begin{figure}
\includegraphics[height=0.48\textwidth, angle=90]{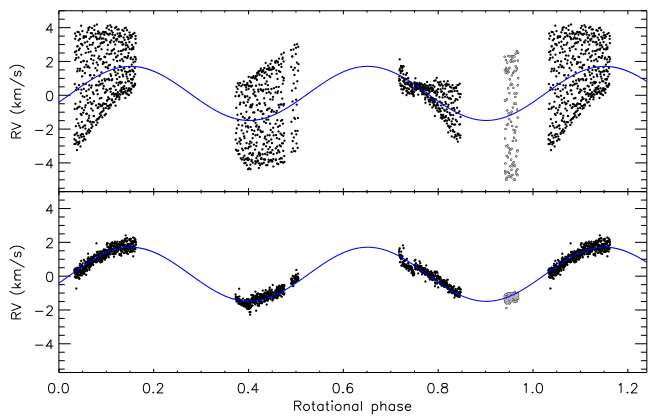}
\caption{\label{fig:ampHI} Radial velocities for HD\,99563 measured in the line 
core of \ha\ for the 2007 observations (black dots) and the 2004 data (grey dots 
at phase 0.94). The modulated pulsation amplitude is superposed on a rotational 
modulation of the average radial velocity (fitted with solid line). Rotation 
phases are relative to Eq.\,\ref{eq:ephem}. The top panel shows measured radial 
velocities, while the bottom panel shows rotational radial velocities 
that remain after removing the pulsational variations by prewhitening 
the {\it pulsation} frequencies (i.e., the pulsational frequency nonuplet) in 
Table\,\ref{tab:frq}. The broad band of data in the top panel is not caused by 
scatter; it is the pulsation compressed in time. Fig.\,\ref{fig:ampHIdetail} shows 
these data expanded in time where the pulsation cycles can be seen.}
\end{figure}

The \halpha\ profile is shown in Fig.\,\ref{fig:haregion}. The insert shows the 
line core, with the region measured indicated with a bar. Radial velocities, 
measured through Gaussian fitting to the line profile cores, are shown in 
Fig.\,\ref{fig:ampHI}, phased with the rotation period. These reveal a double wave 
sinusoidal  variability of the {\it average} 
radial velocities with an amplitude of 
1.6\,\kms\ and with a frequency twice that of the rotation frequency, $2\nu_{\rm 
rot}=0.0079498$\,mHz. Using the frequency analysis tool {\sc PERIOD04} 
\citep{lenzetal05} to improve phases and amplitudes iteratively for the solution 
of \citet{handleretal06} (for fixed frequencies, keeping only the significant 
ones), the $2\nu_{\rm rot}$ modulation is found (Table\,\ref{tab:frq})
to have an amplitude of 1600\,\ms. 
Single or double wave angular modulation of the mean radial velocity of
stars is known from many other elements in spotted stars --
particularly the rare earth elements --  but has not previously been 
detected for hydrogen.

Comparison with telluric lines, constant within 600\,\ms, excludes a non-stellar 
origin (Fig.\,\ref{fig:rotvar}) for the H$\alpha$ rotational radial velocity 
variations. Detailed comparison of averaged spectra in bins of 35 minutes show the 
slow periodicity \changeb{in Fig.\,\ref{fig:rotvar}} dominates in the blue wing and central part of the slightly 
asymmetric \halpha\ core. A blend with a line 
\changeb{of a rare earth element distributed mainly around the stellar poles}  
cannot 
be completely excluded as the source of the radial velocity curve. 
However, we consider 
it unlikely, based on comparison to a synthetic spectrum, that 
such relatively weak lines (compared to the \halpha\ core) 
can result in the amplitude observed. 

The variation in the mean radial velocity of the H$\alpha$ line with rotation is 
shown in Fig.\,\ref{fig:ampHI}, with the pulsation variation present (in the top 
panel) and after prewhitening for the pulsational nonuplet (given in Table~2). 
There are several possible explanations for this rotational variation that we now 
examine. 

\subsubsection{Variations in the mean radial velocity caused by line profile 
variations}

We modelled line profile variations for an oblique dipole pulsation mode 
with a rotational 
inclination of $i = 44^\circ$, pulsation pole obliquity of $\beta = 88^\circ$, an 
equatorial rotational velocity of 30\,km\,s$^{-1}$ and a pulsational amplitude of 
14\,km\,s$^{-1}$  -- values appropriate for HD\,99563. The synthetic line profiles 
were computed using an adapted version of the line profile code implemented in 
\textsc{FAMIAS} (\citealt{zima06}, 2008). The underlying model approximates the 
velocity field by a spherical harmonic, assumes a Gaussian intrinsic profile that 
originates from a two-dimensional atmospheric layer, and is weighted with a 
fourth-order limb darkening law. We adapted the above mentioned code to include 
the effects of oblique pulsation on the line profile variations. 

\begin{figure} 
\begin{center} 
\includegraphics[height=0.47\textheight, angle=0]{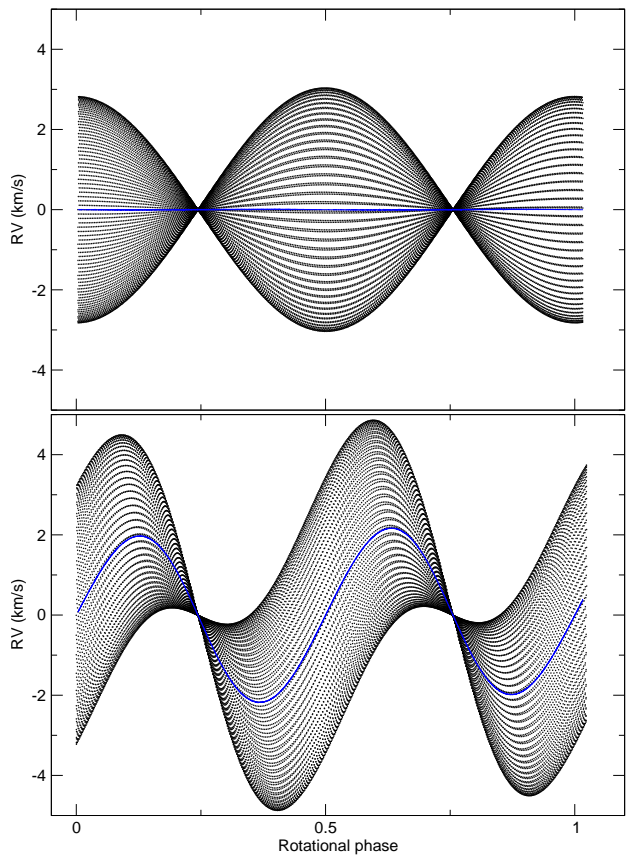}
\caption{\label{fig:zgphp1}Top panel:  Radial velocities over the rotation period 
\changeb{(black dots in electronic version) and its running mean (solid line at
RV=0\,\kms)} for 
model line profile variations of an oblique dipole pulsation mode and uniform 
element distribution with 
$i = 44^\circ$, $\beta = 86^\circ$, $v \sin i = 30$\,km\,s$^{-1}$ and a pulsation 
amplitude of 14\,km\,s$^{-1}$.  The radial velocity amplitude is modulated due to 
the changing aspect angle of the pulsation mode, whereas the mean shows no 
significant variation. \changeb{Abscissa is rotation phase.}
Bottom panel: Same as the top panel, but with a non-uniform 
element distribution that is higher near the magnetic poles, simulating H 
abundance spots. Both the radial velocity amplitude and its mean are modulated due 
to the stellar rotation in the same way as the observations of the H$\alpha$ 
line. As for Fig.\,\ref{fig:ampHI}, the model data here are compressed in time so 
that the pulsations appear only as a broad band. If this figure were expanded in 
time (as in Fig.\,\ref{fig:ampHIdetail} for the H$\alpha$ data), then the 
pulsation cycles would be \changeb{seen. 
The} important features to be noted are the envelope of 
the data, which shows the amplitude modulation, and the mean radial velocity. 
Comparison of the lower panel with the H$\alpha$ data in the upper panel of 
Fig.\,\ref{fig:ampHI} shows the plausibility of our hypothesis that the variation 
in mean radial velocity is caused by enhanced hydrogen spots at the magnetic 
poles. }
\end{center} 
\end{figure}

We computed a first set of synthetic line profiles for a complete rotation cycle, 
assuming uniform element distribution across the stellar surface. 
We then measured the radial velocities for the model data by computing the first 
moment of the line profile.
The top panel of Fig.\,\ref{fig:zgphp1} shows that in this case 
the radial velocity amplitude is modulated whereas the rotational variation of the 
mean radial velocity of the model line is negligible. Therefore, the radial 
velocity variation with the rotation period for H$\alpha$ is not a consequence of 
line profile variations for a simple oblique dipolar pulsation mode.

\subsubsection{An unseen binary companion in a synchronous orbit}

An obvious possible explanation for the radial velocity modulation seen in 
Fig.\,\ref{fig:ampHI} is to hypothesise an unseen binary companion. In this case, 
the orbital period must be equal to the rotation period of HD\,99563, and that is 
reasonable for a 2.91-d rotation period as synchronism is expected for such a 
short orbital period. There is no secondary spectrum seen, thus the single-lined
spectroscopic binary mass function is

\begin{equation}
\frac{(m_2 \sin i)^3}{(m_1 + m_2)^2} = (v_1 \sin i)^3 \frac{P}{2 \pi G} = 1.24 
\times 10^{-6} {\rm M}_{\odot}
\end{equation}

\noindent \changeb{where the primary's projected orbital velocity $v_1 \sin i = 
1600$\,m\,s$^{-1}$ is from Table\,\ref{tab:frq}. 
The component masses $m_1$ and $m_2$ are for the primary (pulsating) component and the
less massive secondary component respectively, $P$ is the orital period (assumed
identical to the primary's rotation period).}
Assuming the orbital plane is the same as HD\,99563's equatorial plane -- a 
reasonable assumption for such a close system -- gives the orbital inclination $i$ to 
be the same as the rotational inclination derived by \citet{handleretal06}, $i = 
44^\circ$. With this inclination the very small mass function shows that $m_1 >> 
m_2$. Taking the mass of HD\,99563 to be that estimated by \citet{handleretal06}, 
$M = 2$\,M$_{\odot}$, gives $m_2 = 0.025$\,M$_{\odot}$, or about 26 Jupiter masses 
with a separation between the components of about 11\,R$_{\odot}$. 

Thus one explanation of the radial velocity variation in H$\alpha$ with rotation 
is a \changeb{close hot Jupiter companion} in a synchronous orbit. The current exoplanet 
listing\footnote{http://exoplanets.org} indicates that this is not unreasonable 
when compared to other known exoplanets. This, if true, is an exciting result, as 
no planet has been discovered orbiting a magnetic Ap star. 

If a planetary companion is the explanation for the H$\alpha$ rotational radial 
velocity variation, then we expect all spectral lines to show the same 
radial velocity curve. Testing this for HD\,99563 is complicated by the 
spotted nature of 
its atmosphere, causing all spectral lines to vary with rotation, but with 
different amplitudes and phases depending on the abundance distributions. For 
example, other elements, such as Mg in Fig.\,\ref{fig:rotvar}, show different mean 
radial velocity curves (also when using telluric lines as reference for the 
wavelength calibration), whereas in the same figure, Fe shows some similarities 
with H$\alpha$. Because the rotational radial velocity amplitudes of various 
elements caused by the spots are comparable to, or higher than, the rotational 
amplitude variation seen in H$\alpha$, a hot Jupiter remains a possible 
explanation, but not our preferred one. 

\begin{figure*}
\includegraphics[height=0.75\textwidth, angle=90]{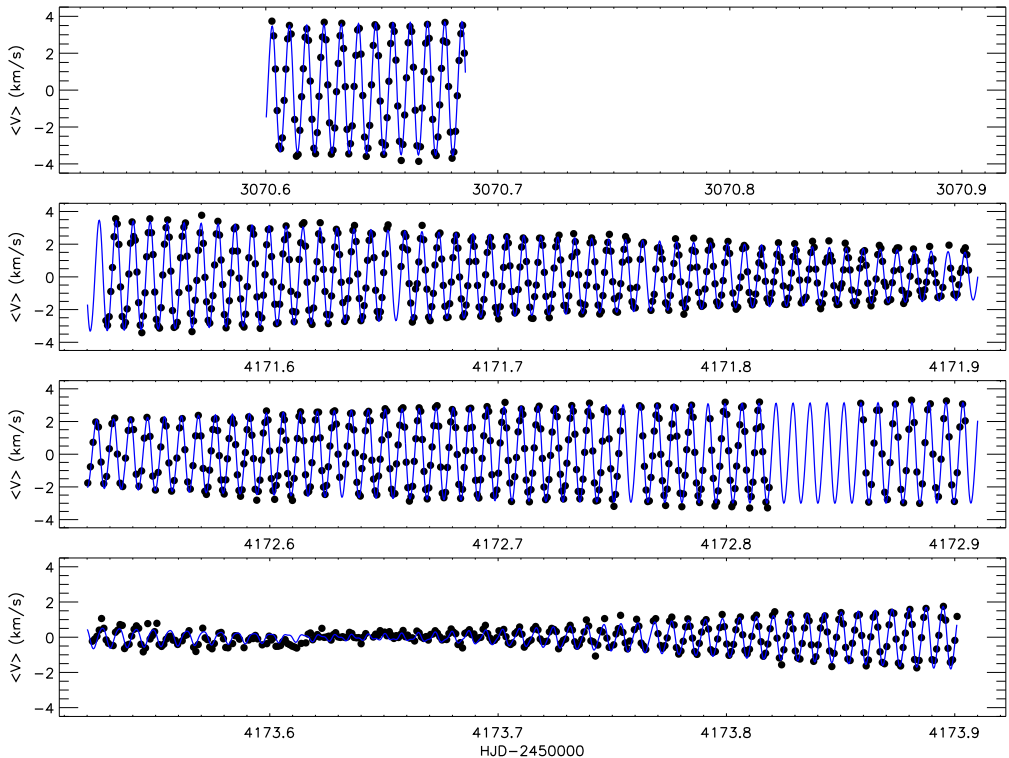}
\caption{\label{fig:ampHIdetail} Radial velocities for HD\,99563 measured in the 
line core of \ha\ for the 2007 observations and the 2004 data (top). Each panel is 
for a separate night. The data were prewhitened for the rotational mean radial 
velocity modulations. Superposed in blue is the frequency solution in 
Table\,\ref{tab:frq} using the distorted oblique dipole nonuplet 
splitting of $\nu$ (1.5576530\,mHz). }
\end{figure*}

\subsubsection{Spots of enhanced H at the magnetic poles}

We suggest this is the first detection of enriched H abundance spots near 
the magnetic poles, as is expected theoretically from helium depletion by 
gravitational settling \citep{balmforthetal01}. 
Such spots give rise to extra absorption superposed on 
the mean \halpha\ core line profile that is variable with the rotation of the 
star. When a hydrogen spot is in the centre of the disk, the radial velocity 
contribution from its extra absorption is zero. Then as it moves away from the 
centre the extra absorption is in the part of the disk that rotates away from us, 
giving rise to a red shift of the line. As the spot reaches the limb, the opposite 
spot is coming into view on the other side of the disk, increasingly cancelling 
out the redshift by a blueshift. At quadrature (or in this case, magnetic 
crossover), both spots cancel out the contribution to the mean profile and the 
radial velocity shift is zero. This is followed by a blue shift. This is what is 
seen in Fig.\,\ref{fig:ampHI}; the phasing is correct; the double-wave is as 
expected. 

To simulate the scenario of enriched element abundance spots near the magnetic 
poles, we followed the same procedure as in Sect.~3.2.2 above, 
but in this case with 
weighting of the local intrinsic line width $W$ with respect to the latitude 
$\theta$ of the displacement field using the following arbitrary function:
\begin{equation}
\label{eq:linewidth}
W(\theta) = W_0 (1 + \cos^2 \theta)
\end{equation}
The resulting shape of the radial velocity curve is displayed in the bottom panel 
of Fig.\,\ref{fig:zgphp1}. The radial velocity amplitude as well as its mean value 
are modulated with an amplitude that is comparable to the observations of the 
H$\alpha$ line.

Thus our hypothesis that the \changeb{observed H$\alpha$ line core's 
radial velocity variation with rotation} 
is caused by theoretically expected hydrogen abundance spots is 
consistent with our simple model. 
Furthermore, and more importantly, we predict that if our hypothesis is 
correct, then hydrogen spots will  be detectable with 
stellar rotation for other Ap stars with strong magnetic fields of similar 
configuration and orientation as HD\,99563. We plan tests of this prediction. 

Of course, even if hydrogen \changeb{was} uniformly distributed, there would be some 
rotation variation in its equivalent width as a result of the cooler spots at the 
magnetic poles caused by \changeb{supression of local convection by the vertical
magnetic flux tubes} and by the non-uniform distribution of other elements -- 
primarily rare earth elements. \changeb{However, in this case the contribution,
if any at all,
from the polar regions is a reduced absorption in the hydrogen lines (for stars
cooler than A5-type stars), which is
quite opposite to our hydrogen spot explanation.}
This does not agree with what is seen in Fig.\,\ref{fig:ampHI}. In 
addition, the general shape and phase of the modulated mean radial velocity curve 
from hydrogen spots must match that of other elements concentrated near the 
stellar poles. Nd is such an example (see the Doppler imaging map in 
Fig.\,\ref{fig:ndDI}) and, except from different amplitude and that the lines 
become double at magnetic crossover, the phases of the modulation are similar
(see Fig.\,\ref{fig:ndpham}).

\subsubsection{Detected frequencies}
\label{sec:hafrqana}

\begin{table} 
\begin{minipage}{81mm} 
\caption{{\normalsize \label{tab:frq}Amplitudes and phases determined from 
a least-squares fit of the listed frequencies to the \halpha\ 
radial velocity series. Column 3 gives the frequencies calculated relative to a 
value of $\nu$ slightly revised from \changeb{that} of \citep{handleretal06}  
(to tie together the 2004 and 2007 data sets) and to 
$\nu_{\rm rot}$. The frequency nonuplet was required to have components 
separated by exactly the rotation frequency. 
A 2\,\cd low frequency periodicity that is believed to be an 
artefact was removed. Errors on $\nu$ and $\nu_{\rm rot}$ 
from \citep{handleretal06}  are 
$\sigma_\nu = 0.000007$\,mHz and 
$\sigma_{\nu_{\rm rot}} = 0.0000001$\,mHz. Prewhitening 
for this solution leaves point-to-point scatter in the residuals of 212\,\ms. 
Gaussian fitting to the profile was used. \changeb{The zero 
point for the time scale 
$t_0 = {\rm HJD}245\,4171.43328$ 
was selected such} 
that the phases of $\nu - \nu_{\rm rot}$ and $\nu + \nu_{\rm rot}$ are equal; i.e. 
this is the time of pulsation amplitude maximum, as given in
Eq.\,\ref{eq:ephem}.}} 
\begin{tabular}[ht!]{@{\,}l@{\,}l@{\,}r@{\,\,}r@{\,\,}r@{\,}r@{\,}} \hline 
Id. &Freq. & Freq.(rev.)&Amplitude& Phase~~~ &$S/N$\\ 
 &(mHz) & (mHz)~~~ &(\ms)~~ & (rad)~~~ & \\ \hline 
$\nu $ & 1.5576530 & 1.5576482 &$ 156\pm31$ &$-2.56\pm0.20 $& 5.1 \\ 
$\nu$$-$$\nu_{\rm rot}$ & 1.5536779 & 1.5536733 &$1726\pm32$ &$-2.17\pm0.02 $& 
54.3 \\ 
$\nu$+$\nu_{\rm rot}$ & 1.5616281 & 1.5616231 &$1303\pm32$ &$-2.18\pm0.02 $& 41.0 
\\ 
$\nu$$-$$2\nu_{\rm rot}$ & & 1.5496984 &$ 121\pm22$ &$-2.11\pm0.18 $& 5.5 \\ 
$\nu$+$2\nu_{\rm rot}$ & & 1.5655980 &$ 80\pm22$ &$ 0.42\pm0.27 $& 3.7 \\ 
$\nu$$-$$3\nu_{\rm rot}$ & 1.5457276 & 1.5457235 &$ 449\pm20$ &$-2.63\pm0.04 $& 
22.2 \\ 
$\nu$+$3\nu_{\rm rot}$ & 1.5695784 & 1.5695729 &$ 105\pm20$ &$ 1.25\pm0.19 $& 5.3 
\\ 
$\nu$$-$$4\nu_{\rm rot}$ & & 1.5417486 &$ 35\pm19$ &$ 1.56\pm0.55 $& 1.8 \\ 
$\nu$+$4\nu_{\rm rot}$ & & 1.5735478 &$ 123\pm19$ &$-1.87\pm0.16 $& 6.4 \\ 
$2\nu $ & 3.1153060 & 3.1152963 &$ 129\pm8\phantom{1}$&$ 0.95\pm0.06 $& 15.9 \\ 
$\nu_{\rm rot} $ & 0.0039749 & &$ 53\pm11$ &$-3.14\pm0.26 $& 4.9 \\ 
$2\nu_{\rm rot} $ & 0.0079498 & &$1600\pm13$ &$-1.90\pm0.01 $& 127.0 \\ 
$2\nu$$-$$2\nu_{\rm rot}$& 3.1073557 & 3.1073465 &$ 72\pm8\phantom{1}$ &$ 
0.48\pm0.11 $& 8.8 \\ 
\hline \hline 
\end{tabular} 
\end{minipage} 
\normalsize 
\end{table} 
 
The frequency resolution of our 2007 data alone, 4.8\,$\mu$Hz, is insufficient to 
improve the period by \citet{handleretal06}. However, by revising the pulsation 
frequency $\nu$ in a test grid within the 1$\sigma$ error given by Handler et al., 
we could also include the 2004 data in the fit. This revision was done by 
recalculating the first seven frequencies by Handler et al. (Column 2 in 
Table\,\ref{tab:frq}) relatively to the values of the trial frequency $\nu$, using 
fixed amplitudes. A small revision (+4.8\,$\mu$Hz) of $\nu$, chosen as the closest 
alias to their value of $\nu$, gave a good fit to the frequency quintuplet and 
residuals of only 245\,\ms\ per observation for all $2004 - 2007$ data and 
10\,\ms\ for the highest peaks in amplitude spectrum of the residuals to the fit. 
The revised frequency is 
henceforth used as $\nu$, but we \changeb{emphasize} that it is a convenient alias out of 
many choices that phases the 2004 data with the 2007 data, 
rather than an improvement in precision to the multisite result by 
\citet{handleretal06}. 
We note that combining the two data sets cannot improve the 
frequency determination more than with the 2007 data alone, due to the 
introduction of several ambiguous aliases that are all good candidate 
frequencies. Yet, the 2004 data supplement the pulsation analysis at other 
rotation phases and were therefore usefully included. 
 
\begin{figure}
\hspace{0.3cm}
\includegraphics[height=0.42\textwidth, angle=90]{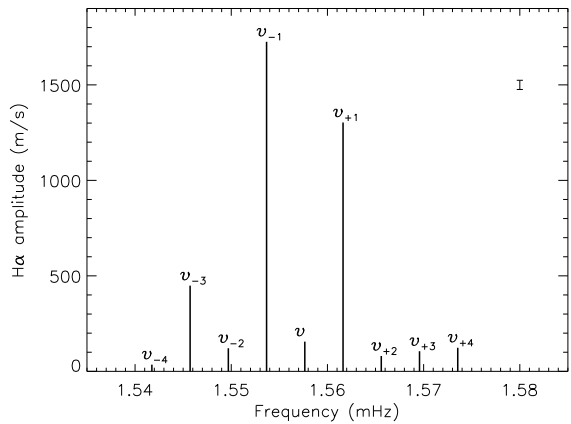}
\caption{\label{fig:frqsol}
The frequency nonuplet detected in the \halpha\ radial velocity series for 
HD\,99563. Mean amplitude error is indicated with a bar (top, right). Note the 
multiplet structure is a result of the distorted dipole pulsator geometry, not 
rotationally perturbed $m$-modes. The frequency splitting is exactly the rotation 
frequency.}
\end{figure}

After prewhitening the full radial velocity series (all $2004-2007$ data) for the 
mean radial velocity modulation (at $2\nu_{\rm rot}$) and the 
first seven significant 
frequencies in the solution by \citet{handleretal06},  we detected excess power in 
the residuals. By including nine terms spaced equally in frequency 
by the rotation frequency, we found all 
but one to be significant; we refer to these nine frequencies as a nonuplet. 
Figs\,\ref{fig:ampHI} and 
\ref{fig:ampHIdetail} show respectively the radial velocity series prewhitened for 
the nonuplet frequencies and for the rotational mean radial velocity modulation. 
The amplitude of hydrogen pulsational radial velocity variations 
ranges from a few hundred m\,s$^{-1}$ to almost 4\,\kms. 
Fig.\,\ref{fig:frqsol} illustrates the nonuplet frequencies, while 
Table\,\ref{tab:frq} (columns $4-5$) gives the corresponding amplitudes and phases 
determined from a linear least-squares fit. 

\subsubsection{Amplitude and phase variations}

\begin{figure}
\hspace{0.2cm}
\includegraphics[width=0.46\textwidth, angle=0]{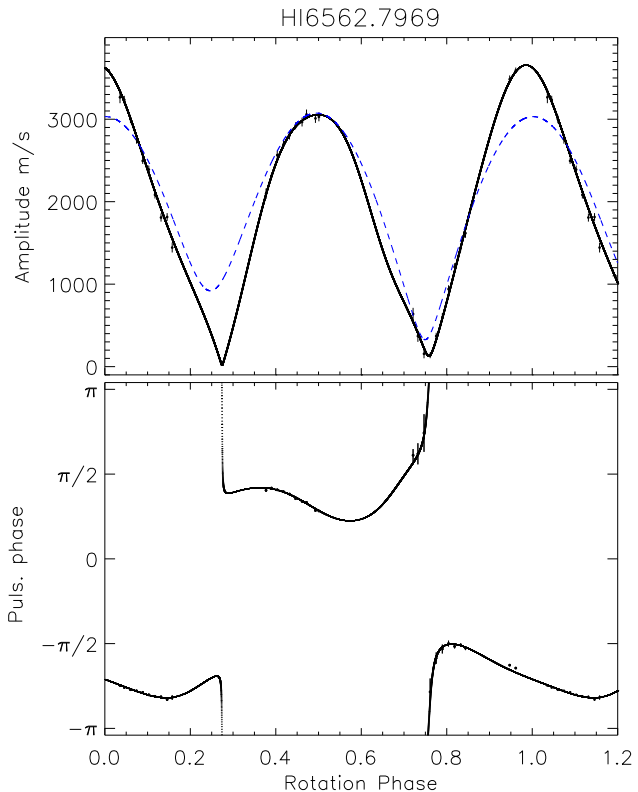}
\caption{\label{fig:haphires}
Amplitude and phase variations from Gaussian fits to the line core of \ha. The 
superposed model in the full line is for a rotationally modulated amplitude 
nonuplet pattern for $\nu$ (see text). For comparison, a pure dipole fit to the 
frequency triplet only is shown with the dashed (blue) line. The lack of spectra 
at $\varphi_{\rm rot}=0.25$ gives the fit too much freedom here.
}
\end{figure}

Amplitude and phase variations were measured with radial velocities from Gaussian 
fits to the line core of \ha\ (Fig.\,\ref{fig:haregion}) in 32 bins of about 1\,hr 
duration. The result is shown in Fig.\,\ref{fig:haphires}. The superposed full 
line is for a rotationally modulated amplitude nonuplet pattern for $\nu$, based 
on amplitudes and phases in Table\,\ref{tab:frq}. The dashed line shows the same, 
but for a pure dipole triplet pattern only. These are based on the general 
($2l+1$) multiplet expression \changeb{in \citet{kurtz92} where} a linear combination of 
axisymmetric spherical harmonics is used to describe the observed angular 
amplitude and phase variations. Amplitudes and phases of 
the multiplet terms were determined from a 
fit to the multiplet frequencies from full 
line profile radial velocity series for that particular line using all spectra. 
For symmetric lines such as \halpha, we used Gaussian fitting. 
Fig.\,\ref{fig:haphires} demonstrates graphically what Table\,\ref{tab:frq} shows 
numerically: The pulsation mode is basically an oblique dipole mode, but 
significant distortion from a simple dipole geometry is required to 
explain the radial velocity amplitude and phase 
modulation with rotation for hydrogen.

\subsection{Neodymium}

\begin{figure}
\hspace{0.5cm}
\includegraphics[width=0.42\textwidth, angle=0]{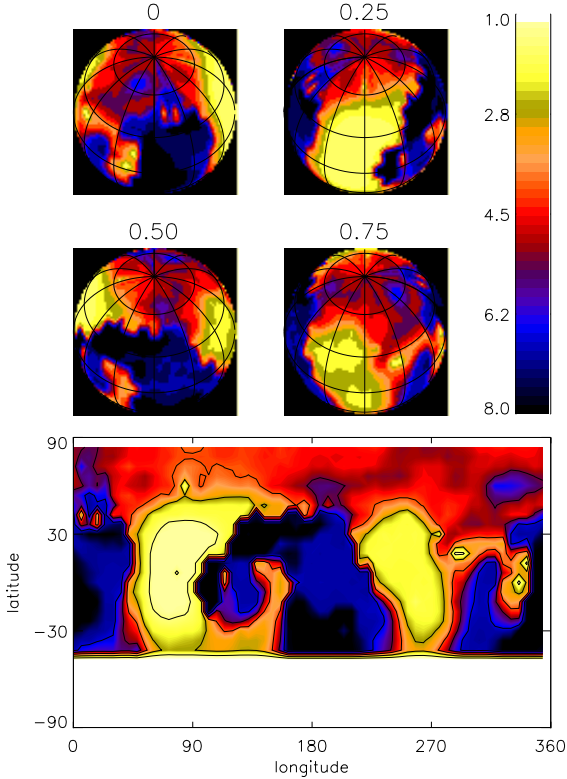}
\caption{\label{fig:ndDI}
Doppler imaging map for Nd\,\textsc{iii}\,6145.07\,\AA.  The abundances are on a 
scale of $\log N_{\rm H} = 12$; the solar abundance of Nd is 1.41 on this scale 
\citep{asplund05}. Thus the highest overabundances of Nd\,\textsc{iii} in the 
spots \changeb{rise to} $10^{6.5}$ times solar at the centres of the two spots associated 
with the 
magnetic poles. These are evident at rotation phases 0.0 and 0.5, 
and they lie near 
the stellar rotational equator (latitude $0^\circ$)  
as a consequence of the magnetic obliquity being 
near $90^\circ$ ($\beta = 88^\circ$). The white colour below latitude 
$-44^\circ$ represents the part of the star that is never visible because of the 
$i = 44^\circ$ rotational inclination.}
\end{figure}

This element is very inhomogeneously distributed on the stellar surface. The 
Doppler imaging map in Fig.\,\ref{fig:ndDI} shows two concentrated spots, 
co-located with the magnetic poles as defined at rotation phases 
0.0 and 0.5 by Handler 
et al. In the analysis of Nd, we will first study the frequency content of the Nd 
lines using the full line profiles, and then study rotational amplitude and phase 
variations for these two spots. For practical purposes (see, e.g., 
Fig.\,\ref{fig:ndpattern} below), we identify these spots as the {\em A} 
(rotational phases $0.2-0.8$) and 
{\em B} (rotational phases $0.7-1.3$) regions.

\subsubsection{Observed line profile variations}

We find significant partial doubling of the Nd lines during the magnetic crossover 
phase, which allows us to study the pulsation of two discrete regions of the 
stellar surface, except for rotation phases near pulsation \changeb{maximums} when only one or 
the other spot is visible. Practically, a direct implication of this is that the 
pulsation amplitude does not go to zero for this element at magnetic quadrature, 
as it does for \halpha. However, as the pulsations in these spots are in 
antiphase, they would cancel out if we could not resolve their individual line 
contributions. This concentration of certain ions, such as those of Nd, in spots 
on HD\,99563 (and some other roAp stars) allows us to study the pulsation 
behaviour of only a part of the visible hemisphere of the star from varying 
aspect, an opportunity provided by no other kind of pulsating star. Ultimately, 
this may allow for full 3-D mapping of the pulsation velocity field, with the 
disentangling of horizontal and vertical velocities -- important for comparison 
with theoretical predictions (e.g., \citealt{saio05}). 

To evaluate the frequency content in the Nd lines, radial velocities were measured 
with Gaussian fits to the symmetric Nd lines Nd\,\textsc{iii}\,5102.41\,\AA\ and 
Nd\,\textsc{iii}\,6327.24\,\AA, using the full line widths, and they were then 
combined. Spectra near rotation phase 0.75 were excluded as that is where the $A$ 
and $B$ pulsation phases become too mixed, or where one of the components strongly 
dominates the other.

\begin{figure}
\includegraphics[height=0.47\textwidth, angle=90]{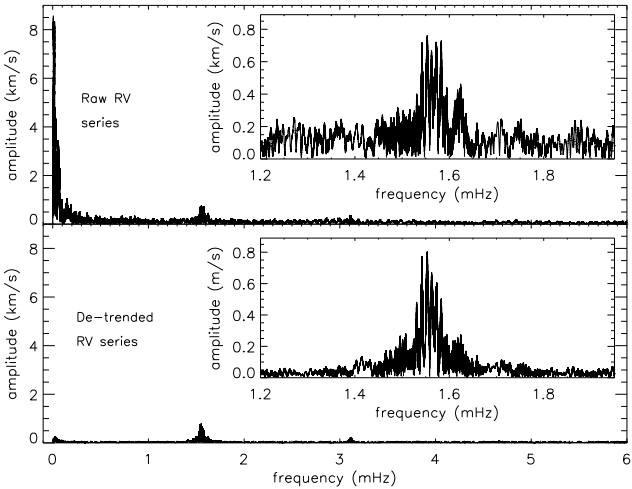}
\caption{\label{fig:nddetrend}
The effect of detrending the `non-binned' radial velocity curve for 
Nd\,\textsc{iii}\,5102.41\,\AA. Top panel shows the amplitude spectrum of the raw 
radial velocity series (with zoom-in panel of the region with the nonuplet 
pulsational frequency content), while the bottom panel shows the amplitude 
spectrum for the detrended radial velocity series. }
\end{figure}
\begin{figure}
\includegraphics[width=0.241\textwidth, angle=0]{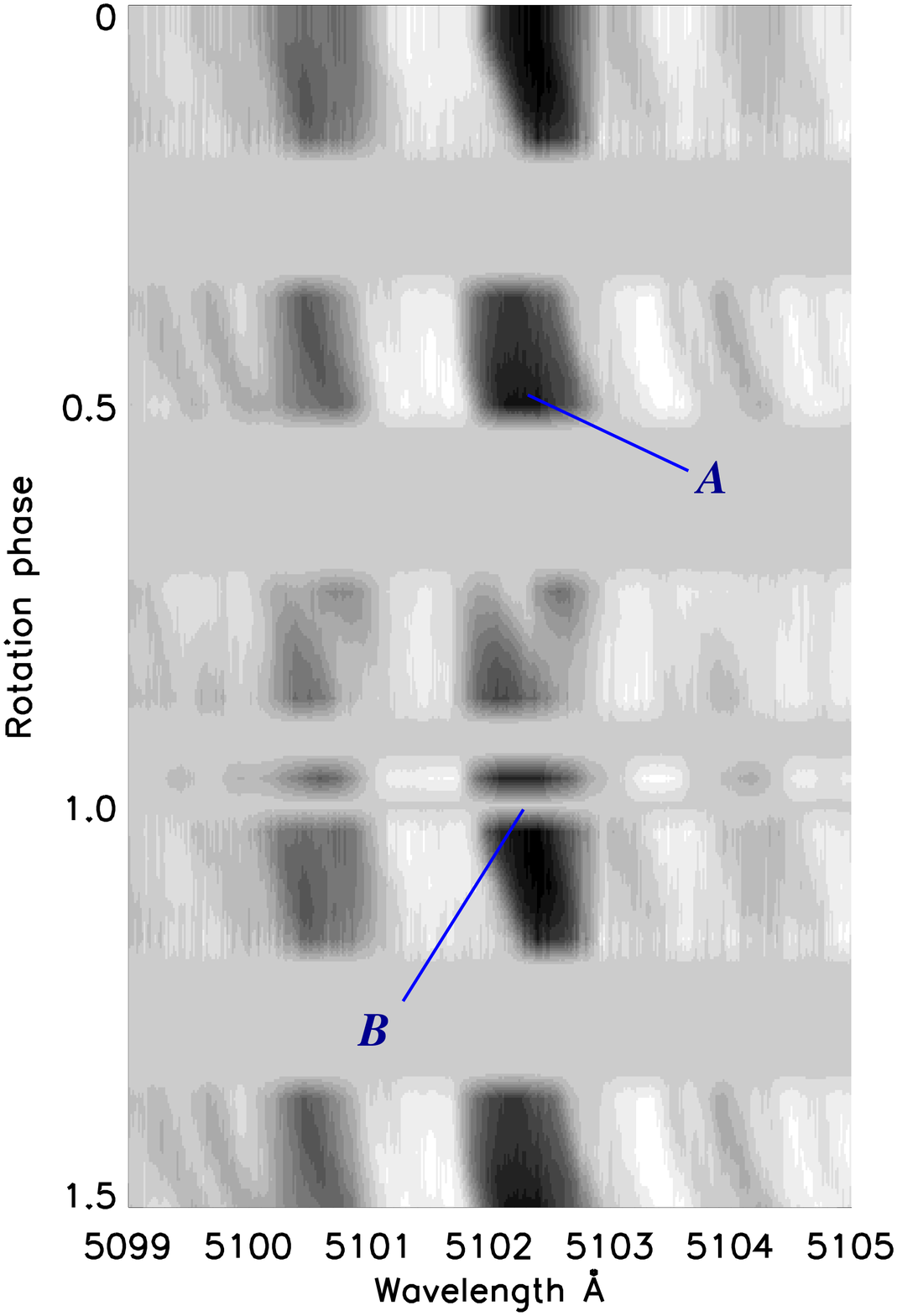}
\hspace{-.34cm}
\includegraphics[width=0.241\textwidth, angle=0]{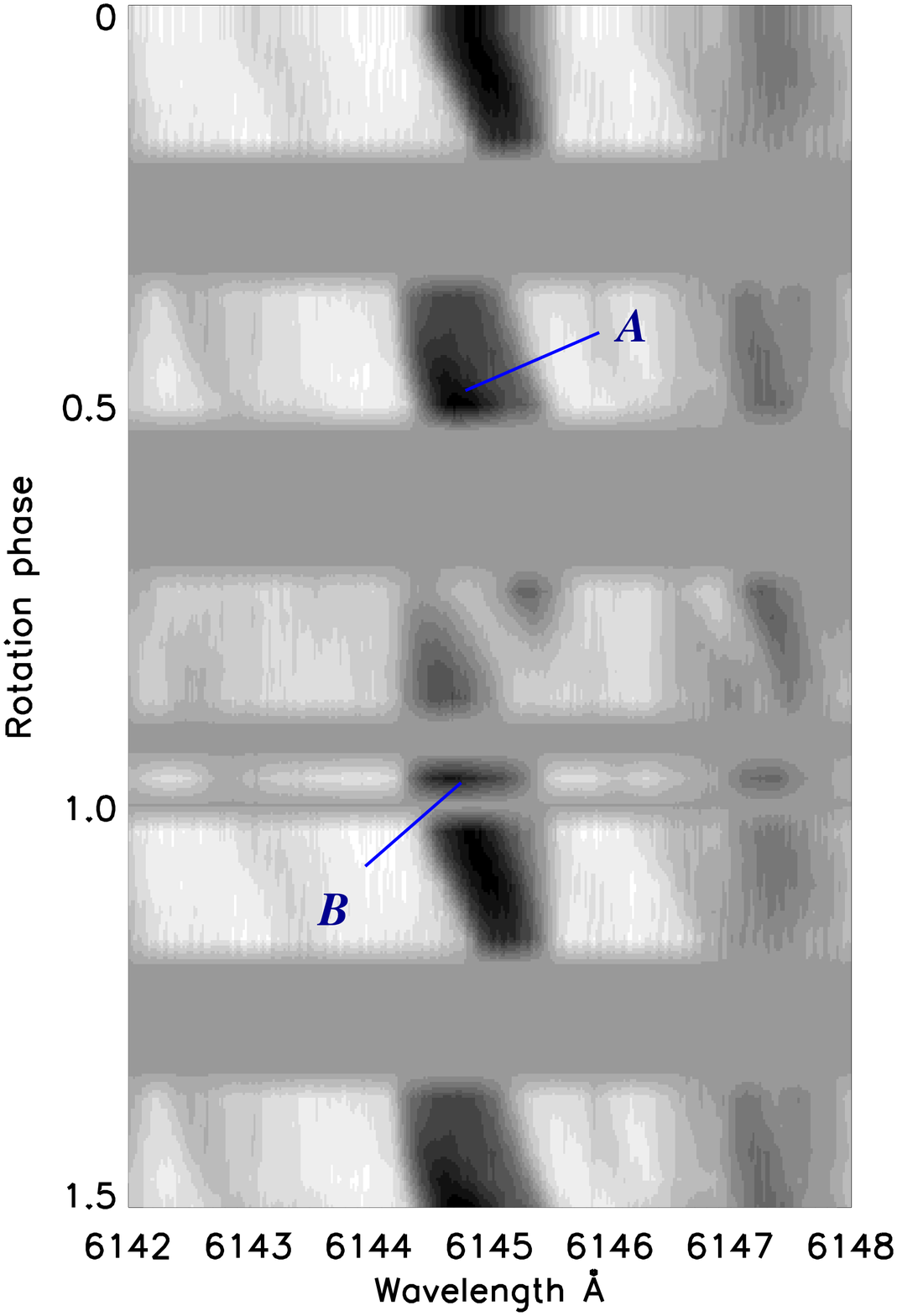}
 \caption{\label{fig:ndpattern}
 Line region in the vicinities of Nd\,\textsc{iii}\,5102.41\,\AA\ (left) and 
Nd\,\textsc{iii}\,6145.07\,\AA\ (right) shown for all spectra sorted according to 
rotation phase of HD\,99563. The spectra are normalised to unity in the continua 
(white colour) and absorption features appear dark or black. Four phase intervals 
not covered by our high speed spectroscopic observations appear with a grey-scale 
corresponding to 0.95. Two line features relate to the dominant surface patches of 
high Nd concentration. These are referred to as the {\em A} 
(rotational phases $0.2-0.8$) and 
{\em B} (rotational phases $0.7-1.3$) regions. }
\end{figure}

The radial velocity series for each spot were each detrended with a 
linear or polynomial fit prior to the frequency analysis (an example of this 
detrending is shown in Fig.\,\ref{fig:nddetrend}). That is justified as
the average radial velocities of the two spots shift quasi-linearly with rotation 
phase, as seen in Fig.\,\ref{fig:ndpattern}. 
A few low frequencies in the 
range $2-5$\,\cd were also included in the least-squares fit to get 
better error estimates. For the fixed nonuplet frequency solution, as for 
 \halpha\ 
given in Table\,\ref{tab:frq}, we find all components to be significant, including 
$\nu-4\nu_{\rm rot}$. The residuals show a flat amplitude spectrum with no 
indication of other pulsation frequencies. Again, as for hydrogen, the many 
multiplet elements indicate a significantly distorted dipole pulsation. 

\begin{table} 
\begin{minipage}{81mm} 
\caption{\normalsize \label{tab:ndfrq}Amplitudes and phases fitted to combined 
radial velocities of two neodymium lines. 
The zero point from \halpha\ in Eq.\,\ref{eq:ephem} was used and the
different phases for the two main sidelobes $\nu$$\pm$$\nu_{\rm rot}$  
correspond to different zero points for times of maximum amplitude modulation
for H and Nd. In fact, 
this difference in time is 0.19464\,d with the Nd maximum coming before the
H maximum. \changeb{This occurs 0.067 earlier in rotation phase,} 
or 24 degrees earlier in
rotational longitude -- in good agreement with the Nd abundance map,
Fig.\,\ref{fig:ndDI}.}
\hspace{.3cm} 
\begin{tabular}[ht!]{@{\,}l@{\,}l@{\,}r@{\,\,}r@{\,\,}r@{\,}r@{\,}} 
\hline 
Id. &Freq. &Amplitude& Phase~~~ &$S/N$\\ 
 &(mHz) &(\ms)~~ & (rad)~~~ & \\ 
\hline 
$\nu $ & 1.5576482& $ 328\pm64$ & $-0.28\pm0.20$ & 5.1 \\ 
$\nu$$-$$\nu_{\rm rot}$ & 1.5536733& $1131\pm71$ & $-2.87\pm0.06$ & 15.8 \\ 
$\nu$+$\nu_{\rm rot}$ & 1.5616231& $ 531\pm72$ & $ 2.57\pm0.13$ & 7.4 \\ 
$\nu$$-$$2\nu_{\rm rot}$ & 1.5496984& $ 445\pm50$ & $ 1.09\pm0.11$ & 8.9 \\ 
$\nu$+$2\nu_{\rm rot}$ & 1.5655980& $ 456\pm50$ & $-1.54\pm0.11$ & 9.1 \\ 
$\nu$$-$$3\nu_{\rm rot}$ & 1.5457235& $ 305\pm43$ & $-1.19\pm0.14$ & 7.2 \\ 
$\nu$+$3\nu_{\rm rot}$ & 1.5695729& $ 192\pm43$ & $-0.68\pm0.22$ & 4.5 \\ 
$\nu$$-$$4\nu_{\rm rot}$ & 1.5417486& $ 230\pm43$ & $-3.04\pm0.19$ & 5.3 \\ 
$\nu$+$4\nu_{\rm rot}$ & 1.5735478& $ 171\pm43$ & $ 2.02\pm0.25$ & 4.0 \\ 
$2\nu $ & 3.1152963& $ 77\pm18$ & $ 0.68\pm0.23$ & 4.4 \\ 
$\nu_{\rm rot} $ & 0.0039749& $ 134\pm31$ & $-0.63\pm0.26$ & 4.4 \\ 
$2\nu_{\rm rot} $ & 0.0079498& $ 141\pm31$ & $ 1.69\pm0.23$ & 4.5 \\ 
$2\nu$$-$$2\nu_{\rm rot}$& 3.1073465& $ 158\pm18$ & $ 1.13\pm0.11$ & 8.9 \\ 
\hline \hline 
\end{tabular} 
\end{minipage} 
\normalsize 
\end{table} 

\subsubsection{Pulsation properties from abundance patches}

We identified the `spot' line profile patterns in Fig.\,\ref{fig:ndpattern} as 
{\em A} (rotational phases $0.2-0.8$) and 
{\em B} (rotational phases $0.7-1.3$). All spectra were 
re-measured by fitting Gaussian profiles to the lines, while tracing the mean 
movement of the $A$ and $B$ features individually with the rotation phase. 
For time intervals of 1\,hr 
the pulsation amplitude and phase were measured for fixed $\nu = 1.5576482$\,mHz, 
the actual pulsation frequency of the distorted dipole mode.

\begin{figure}
\includegraphics[height=0.48\textwidth, angle=90]{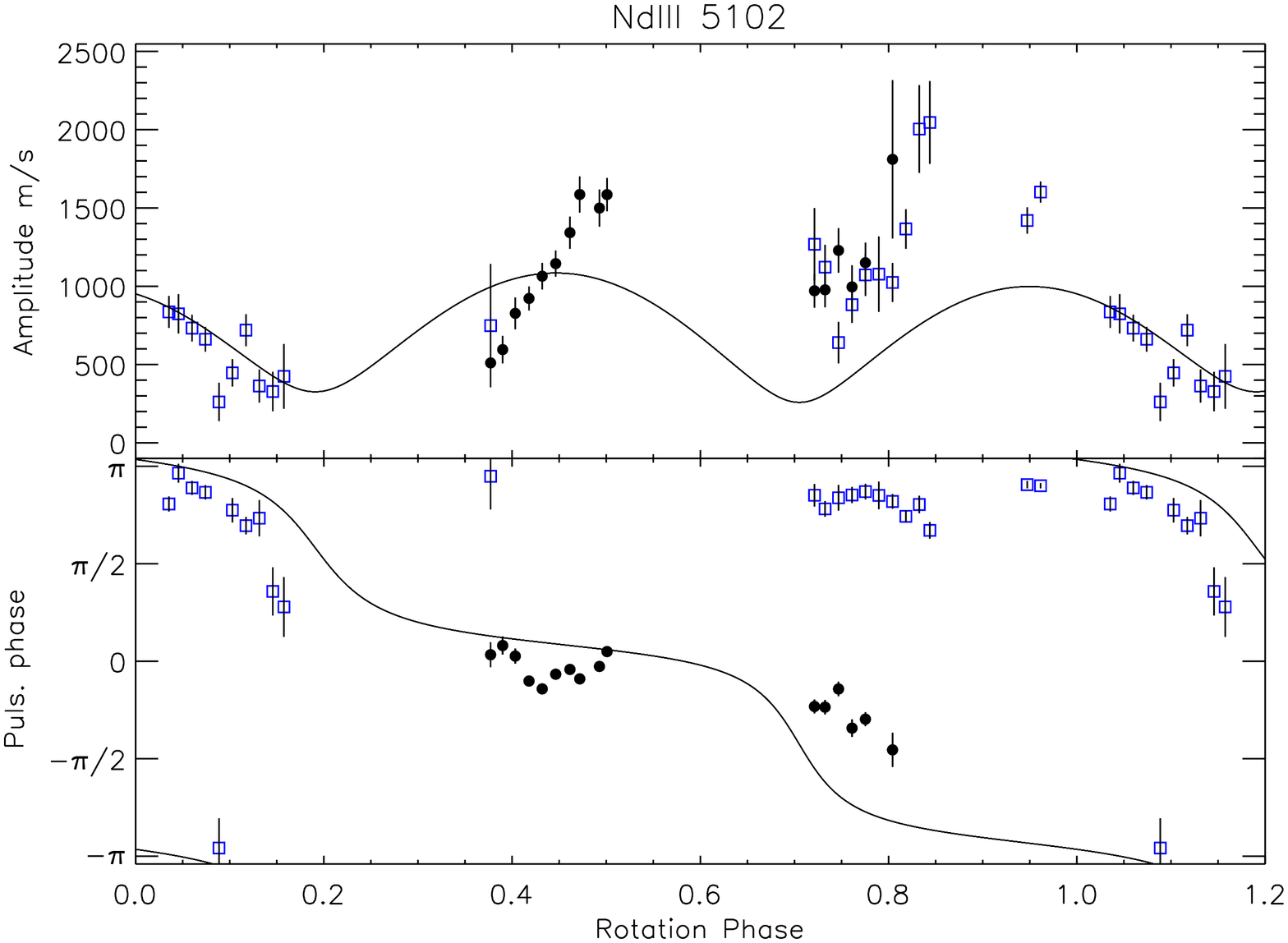}
\vspace{0.2cm}
 \includegraphics[height=0.48\textwidth, angle=90]{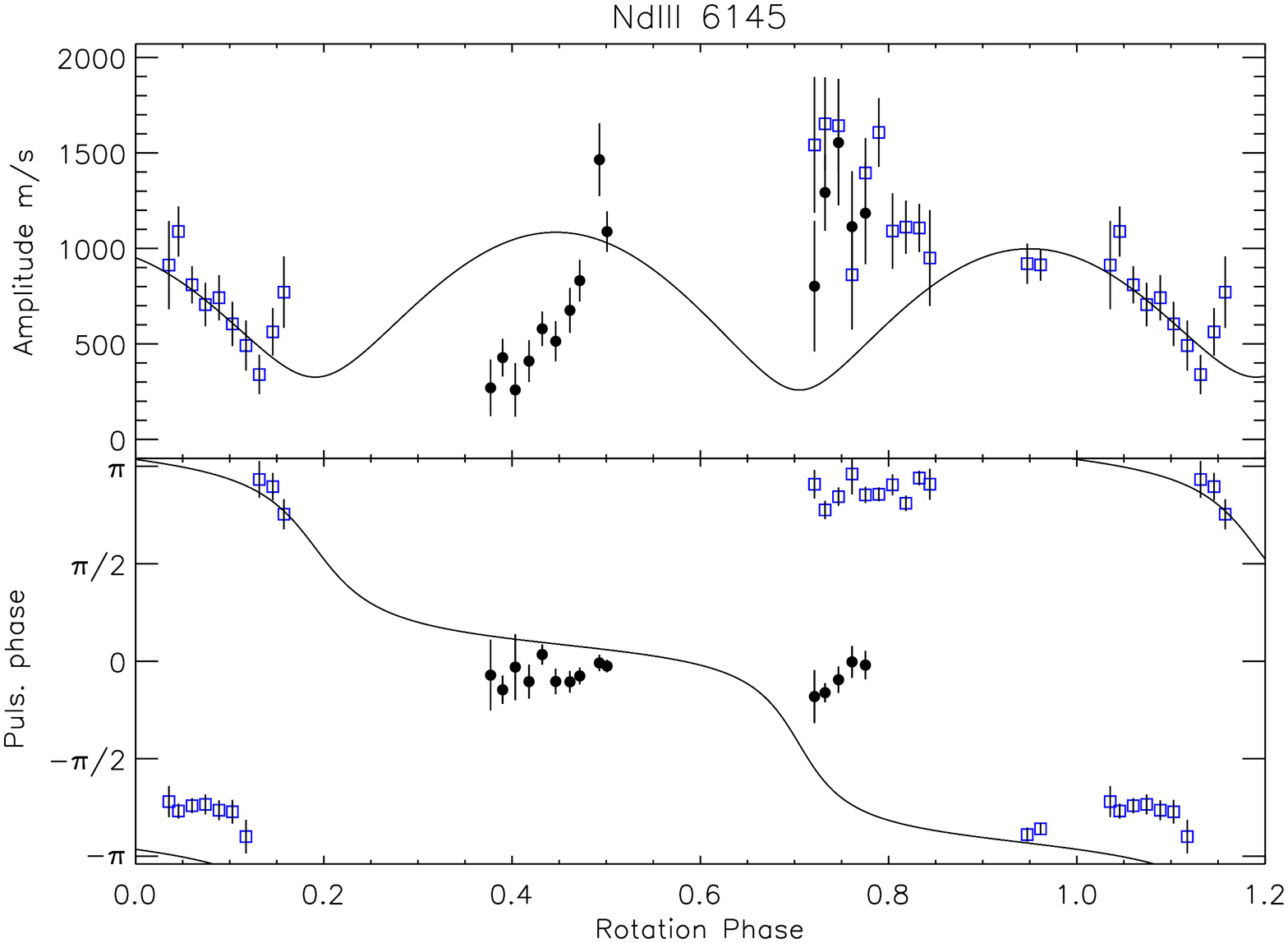}
 \caption{\label{fig:ndpham}
Amplitude and phase variations for Features $A$ (closed circles) and 
$B$ (open boxes) in Fig.\,10 
for Nd\,\textsc{iii}\,5102.41\,\AA\ (top) and Nd\,\textsc{iii}\,6145.07\,\AA\ 
(bottom), measured with Gaussian profile fits. A superposed triplet dipole pattern 
is shown for comparison with a solid line. 
The group of measurements near the magnetic crossover 
phase $\varphi_{\rm rot}=0.75$ occur during the double-lined phase and show a 
phase shift near $\pi$ rad between the two poles, as expected for a basically 
dipolar pulsation mode. }
\end{figure}

The two panels in Fig.\,\ref{fig:ndpham} show the resulting amplitude and phase 
variations with rotation phase for the two lines Nd\,\textsc{iii}\,5102.41\,\AA\ 
and Nd\,\textsc{iii}\,6145.07\,\AA. Measurements for each spot are indicated with 
different symbols -- open boxes for feature $B$ and closed circles for feature 
$A$. Superposed 
on the figure is a dipole model, 
constructed using a triplet frequency fit to the full 
Nd series. Several details are notable in the figure: the pulsation amplitude 
never reaches zero for any of the spots even when at the stellar limb; the 
modulated amplitude maximum for spot $A$ near rotation phase 0.5 (the spectra only 
cover the ingress part), is sharper, steeper and shifted in rotation phase 
with respect to what is expected for a pure dipole model. 
This \changeb{is the result of concentration of Nd\,\textsc{iii} 
towards the magnetic and pulsation poles} (see 
Fig.\,\ref{fig:ndDI}). 
During quadrature at rotation phase 0.75, it is seen that the 
pulsation amplitudes for both spots are similar, and about 2/3 of the maximum of 
the amplitude modulation, a consequence of the projection factor (see below). 
The increased scatter at this phase is due to line 
blending between the double-line components, and weaker absorption lines, as the 
Nd abundance is less overabundant over 
the visible hemisphere of the star at this phase. Note 
however, how clearly the pulsation phases divide for the two spots, 
separated by $\sim \pi$ rad, 
clearly indicating the global $l=1$ dipole pulsation. 

With both spots at the limb of the star, and with the ability to study them 
separately, it is interesting to ask whether we can see any evidence of a 
horizontal component to the pulsations. For example, if the pulsation were purely 
horizontal, rather than vertical (radial), we would expect to see maximum 
pulsation amplitude at this rotational phase \changeb{(0.75)}
for the Nd\,\textsc{iii} lines. On 
the other hand, because the Nd\,\textsc{iii} spots are large, there is a component 
of pulsation in the direction of the line-of-sight 
for purely vertical pulsation. A 
simple integral over the half of the visible hemisphere in which a spot is seen 
(see, e.g., \citealt{parsons72} for the simpler case of Cepheid pulsation) indicates 
that for purely vertical pulsation velocities 
we will still expect to see an amplitude about 
2/3 that seen at pulsation maximum when the spot is closest to the line-of-sight. 
This is what we do see in Fig.\,\ref{fig:ndpham}, hence there is no obvious 
evidence of a horizontal component to the velocity field. We will model this in 
more detail in a future publication to put limits on the horizontal component. 
\changeb{However, as mentioned in Sect.\,1.3 \citet{saio05} shows the horizontal 
motion is expected to be zero at the magnetic poles. Thus the best strategy for 
detecting horizontal motion is through chemical spots located away from the 
 poles.}

\begin{figure*}
\includegraphics[width=0.94\textwidth, angle=0]{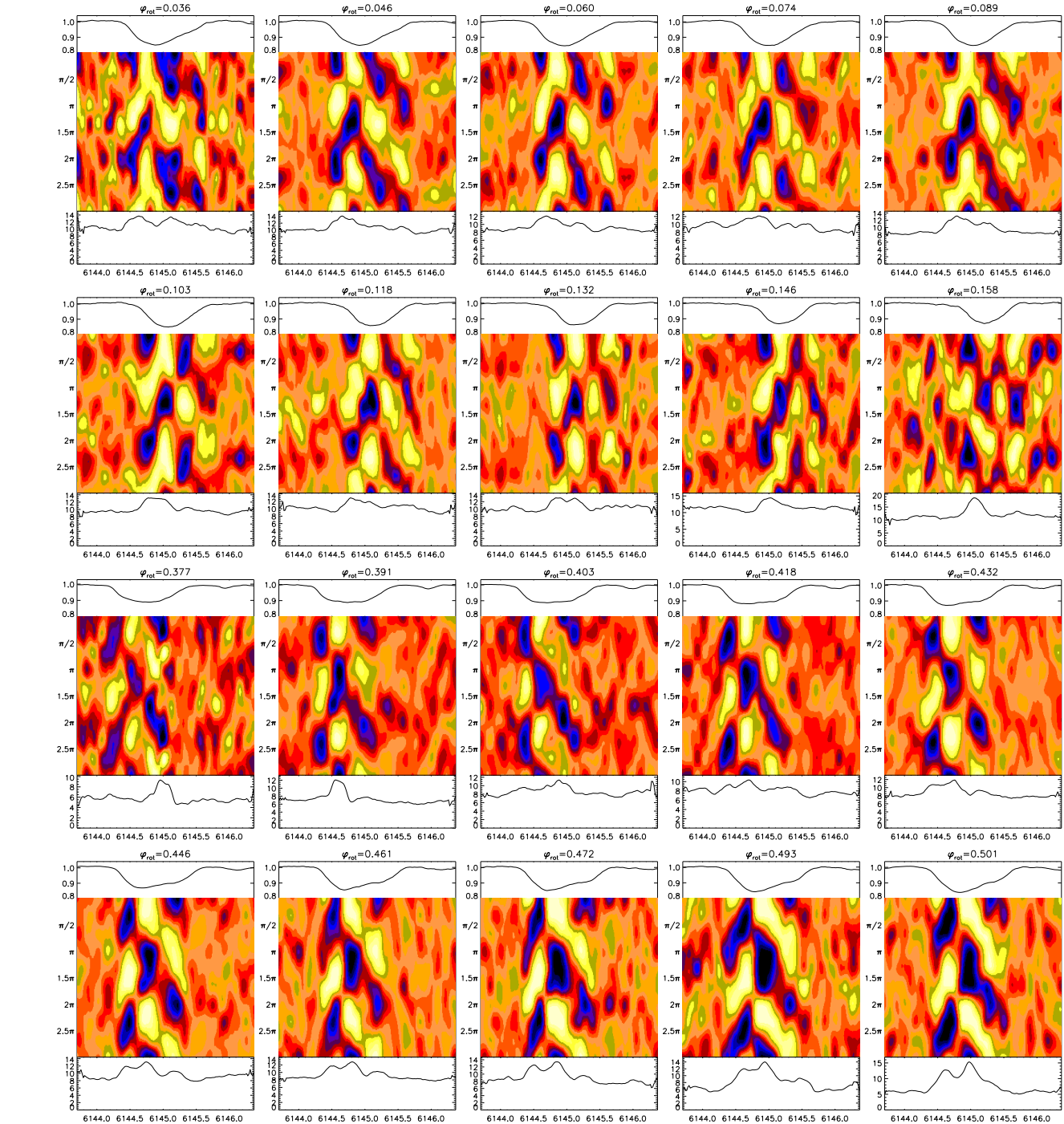}
\caption{\label{fig:ndlpv1}
Residual line profile variations in Nd\,\textsc{iii}\,6145.07\,\AA\ for the 32 
time series bins, with rotation phase increasing to the right and downwards, each 
subtracted from the average spectrum (shown on top of each panel) of that bin. 
Pulsation phase increases downwards in the middle panels. 
The standard deviation of each 
bin series is shown in the bottom of the panels. 
The colour codes are $-0.01$ (black) to $+0.01$ (white) in normalised intensity.}
\end{figure*}

\setcounter{figure}{11}

\begin{figure*}
\includegraphics[width=0.94\textwidth, angle=0]{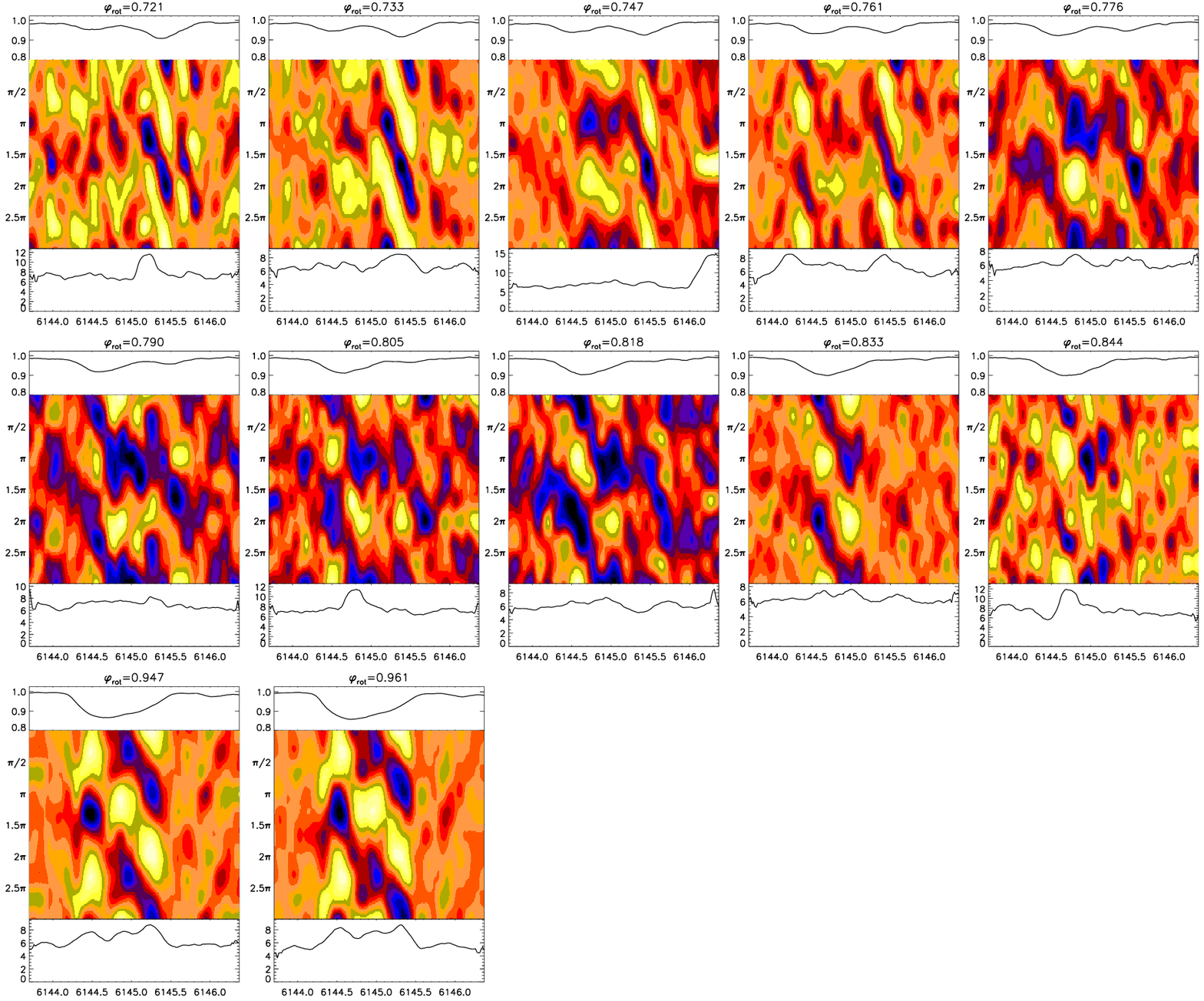}
\caption{\label{fig:ndlpv2}
Fig.\,12 continued. Note the clear line doubling for the rotation phases in the 
top row, observed near the time of magnetic quadrature.}
\end{figure*}

Figure\,\ref{fig:ndlpv1}  shows the line 
profile variability in Nd\,\textsc{iii}\,6145.07\,\AA\  for 
all \changeb{32 1-hr bins  
that we divided} the spectra into. 
Each sub-panel shows the average spectrum at this 
rotation phase, the residual variability for the spectra phased with the pulsation 
phase progressing downwards. The lower insert of each panel is the standard 
deviation for the individual series of spectra, pixel-by-pixel. Flux is in all 
cases in units of normalised flux of the stellar continuum. The figures show a 
$\pi$\,rad pulsation phase reversal from rotation phase $\varphi_{\rm rot}=0.0$ to 
0.5 (seen as a colour inversion) that can be seen to progress with the stellar 
rotation. 

Recently, \cite{shibahashietal08}  have interpreted blue-to-red running 
wave features in the line profile variability in the roAp star $\gamma$\,Equ as a 
manifestation of an upper atmospheric shock wave. Similar blue-to-red features are 
also seen in the line profile variations of other roAp stars (see, e.g., 
\citealt{kochukhovetal07}).  However, up to now observations have generally been 
made for short time spans of typically 2\,h and do not cover different rotational 
phases. An important exception to this is the pioneering study of line profile 
variability in the roAp star HR\,3831 by Kochukhov (2006; see his \changeb{figures
2} and 3) 
which shows similarities in the behaviour to that seen in Fig.\,\ref{fig:ndlpv1} 
for HD\,99563.

We see in Fig.\,\ref{fig:ndlpv1} the line profile variations of a 
spectral line simply shifting in wavelength without a change of shape at phases 
near pulsation maximum when the pulsation axis is closest to the line-of-sight, as 
is expected for a dipole pulsation mode at this aspect. This is clear in the top 
row of the figure at phase, e.g., $\varphi_{\rm rot}=0.046$ and other phases near 
it, and again in the bottom right panel of the figure at phase $\varphi_{\rm 
rot}=0.501$. Yet remarkably at magnetic quadrature when the Nd\,\textsc{iii} line 
is doubled and we can examine the individual components seen at an aspect angle 
near $90^\circ$, there is a clear blue-to-red running wave component visible. Note 
particularly in Fig.\,\ref{fig:ndlpv1} (continued) the upper row of panels and, 
e.g., $\varphi_{\rm rot}=0.733$. This same behaviour was reported first for 
HR\,3831 by \cite{kochukhov06}  (who also reported red-to-blue features which we 
possibly see for HD\,99563 at some rotation phases, e.g. 0.146 in Fig.\,12) and 
is clearly important for our understanding of these 
running wave features seen both here and in other roAp stars.

Further interpretation of these features for Nd, as well as for other 
lines and for other stars, requires modelling of the pulsations, which is work now 
in progress. Whatever the cause of the blue-to-red running wave line profile 
variations -- whether shock waves as proposed and modelled by 
\cite{shibahashietal08}, or some other cause -- 
\changeb{the fact that they are aspect dependent 
with rotation in HD\,99563 and HR\,3831 is an important characteristic that needs 
to be explained.}

\subsection{Europium}

\begin{figure} \includegraphics[width=0.42\textwidth, angle=0]{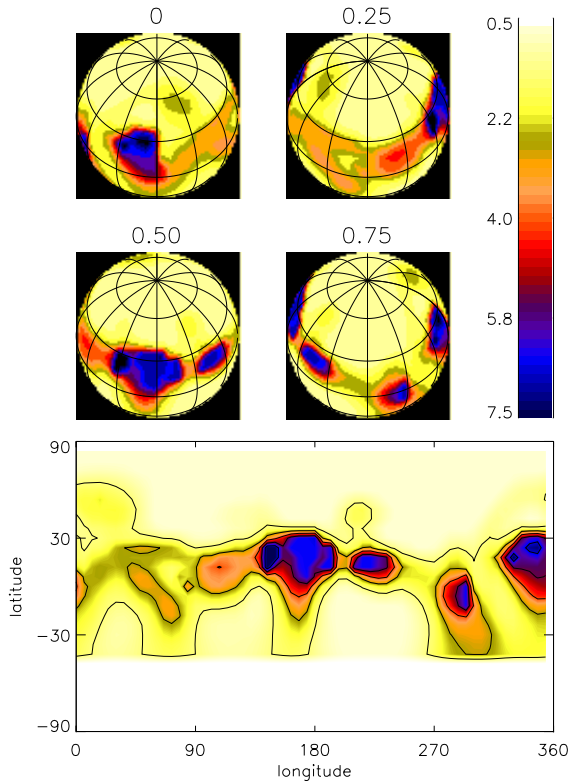} 
\caption{\label{fig:euDI} Doppler imaging map for Eu\,\textsc{ii}\,6645.06\,\AA.  
The abundances are on a scale of $\log N_H = 12$; the solar abundance of Eu is 
0.48 on this scale (Asplund et al. 2005). Thus the highest \changeb{overabundance} of 
Eu\,\textsc{ii} in the spots rises to $10^{7}$ times solar at the centres of the 
three spots, two of which are offset from the 
magnetic and pulsation poles, and the third which is \changeb{isolated at} longitude 
$290^\circ$. The white colour below latitude 
$-44^\circ$ represents the part of the star that is never visible because of the 
$i = 44^\circ$ rotational inclination.}
\end{figure}

This element shows a highly inhomogeneous surface distribution in the Doppler 
imaging map seen in Fig.\,\ref{fig:euDI}. Two circumpolar patches are viewed most 
favourably near the line-of-sight around $\varphi_{\rm rot}=0.45$ and 0.95; the 
third spot is \changeb{rather concentrated and appears nearest the} line-of-sight at 
$\varphi_{\rm rot}=0.65$.

\subsubsection{Observed line profile variations}

\begin{figure}
\includegraphics[width=0.250\textwidth, angle=0]{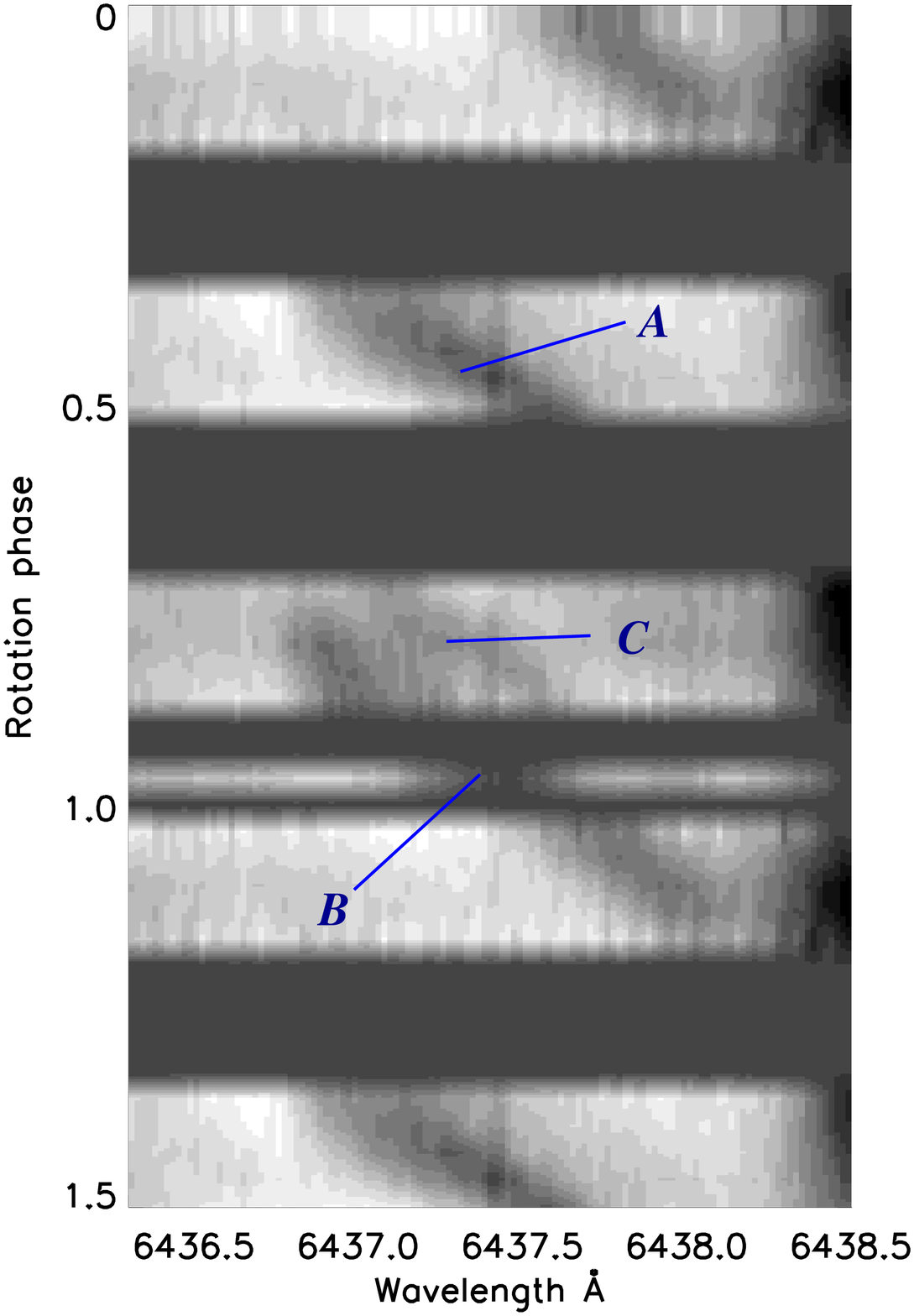}
\hspace{-.4cm}
\includegraphics[width=0.235\textwidth, angle=0]{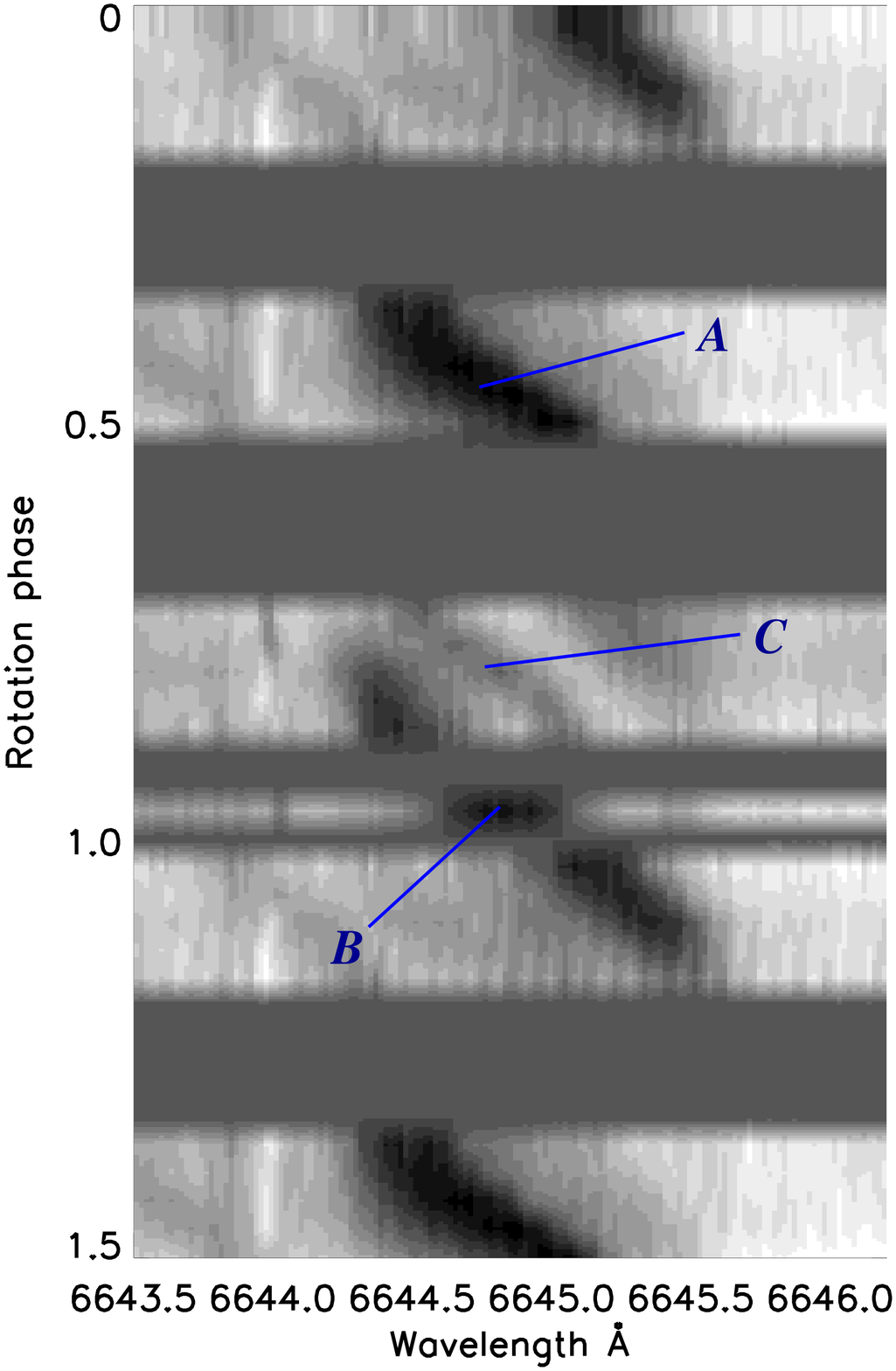}
 \caption{\label{fig:eupattern} Line region in the vicinities of 
Eu\,\textsc{ii}\,6437.64\,\AA\ (left) and Eu\,\textsc{ii}\,6645.06\,\AA\ (right) 
shown for all spectra sorted according to rotation phase of HD\,99563. Absorption 
features appear darker and in black. Four phase intervals were not covered by fast 
spectroscopy and are shown as smooth dark gray bands. 
Two line features relate to dominant surface patches of high Eu 
concentration. These are referred to as {\em A} 
(rotational phases $0.2-0.8$) and {\em B} 
($0.7-1.3$). A third, much more shallow feature is seen at rotational 
phases $0.7-0.8$. }
\end{figure}

In the sequence of spectra in Fig.\,\ref{fig:eupattern}, the two strongest Eu 
lines Eu\,\textsc{ii}\,6437.64\,\AA\ and Eu\,\textsc{ii}\,6645.06\,\AA\ show 
three moving features which are identified in the figure as $A$, $B$ and $C$.

Table\,\ref{tab:frq2} shows the results of frequency analyses of three elements, 
Nd, Eu and Ca. The radial velocities were measured with Gaussian fitting or the 
CoG procedure (for Ca) to the full line profiles in all spectra. To avoid spectra 
with mixed contributions, some $\sim$40 spectra around the magnetic crossover 
phase ($\varphi=0.75$) were not used.  Detrending for low frequencies was 
performed 
with first and second order polynomial fitting to spectra dominated by the $A$ or 
$B$ features separately (third and fourth order polynomials for Eu and Ca) and up 
to 3 low frequencies were prewhitened. The results indicate that Nd requires a 
nonuplet solution and thus has a high distortion to the oblique dipole modulation, 
while Eu has little power left after a triplet fit. Ca \changeb{may, however,} require at 
least a quintuplet pattern, but as it has a very low amplitude of a few hundred 
\kms\ \citep{elkinetal05}, the frequencies are not significant when fitting the 
full nonuplet solution. These results are consistent with the abundance 
distributions of these elements. The small $A$ and $B$ spots of Eu 
do not sample much of the visible hemisphere of the star, 
except near the poles, so they are fitted well by a simple dipole model. 
That means that the distortions to the dipolar pulsation that give rise to the 
other members of the nonuplet for Nd come from regions further from the poles than 
the Eu samples. These differences may eventually be exploited for more detailed 
models of the pulsation velocity field. The high 5\,km\,s$^{-1}$ 
amplitude seen for Eu is 
consistent with its concentration towards the poles, and may also indicate that it 
is radiatively stratified to high levels in the atmosphere where the pulsation 
amplitude is higher. 
The phase differences (relative to Eq.\,\ref{eq:ephem}) of the two main multiplet 
sidelobes of Eu in Table\,\ref{tab:frq2} are, such as
shown for Nd, at least partly due to different surface distributions than H.

\subsubsection{Pulsation properties from abundance patches}

\begin{figure}
\includegraphics[height=0.44\textwidth, angle=90]{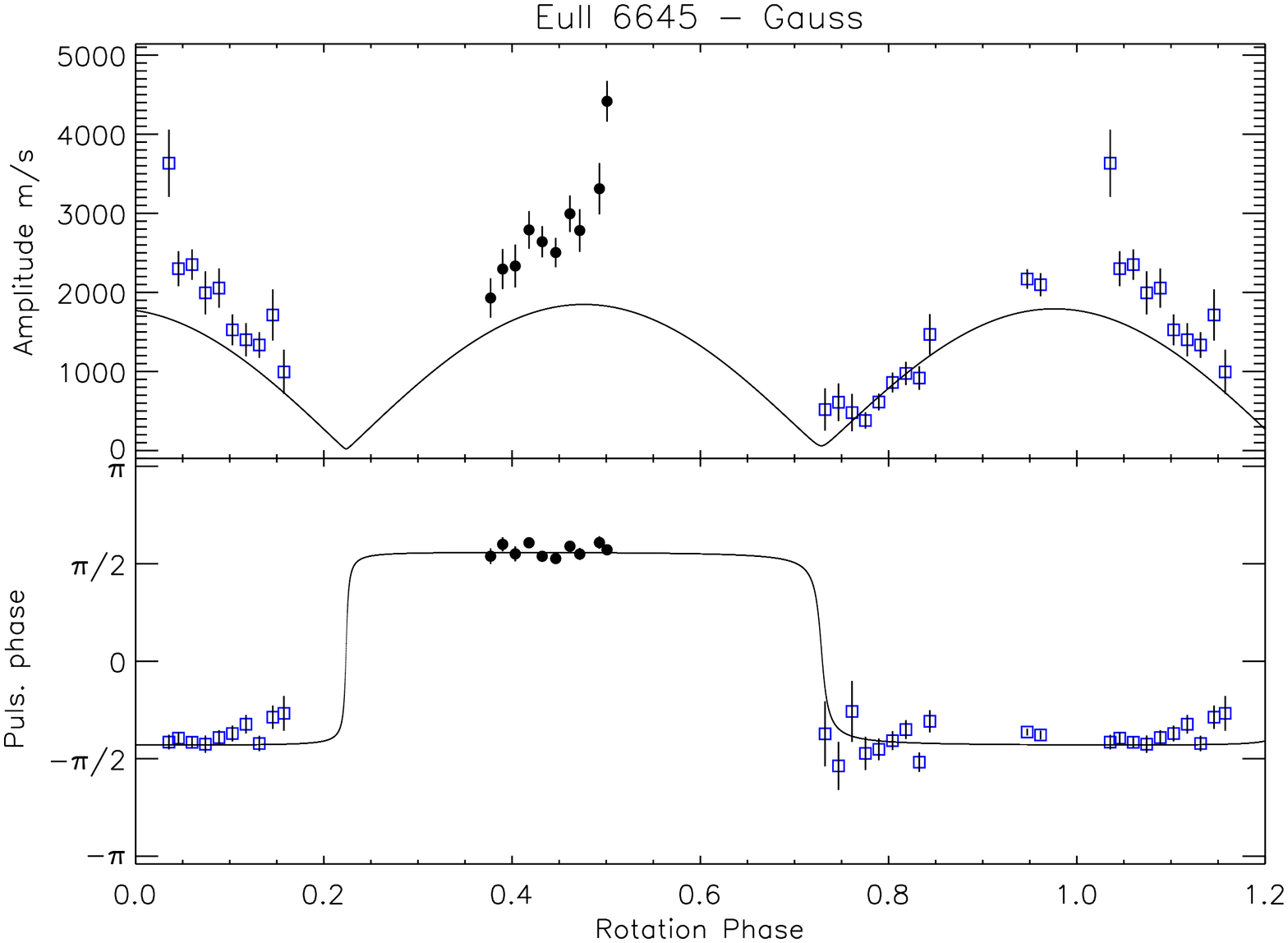}
\vspace{0.2cm}
\includegraphics[height=0.44\textwidth, angle=90]{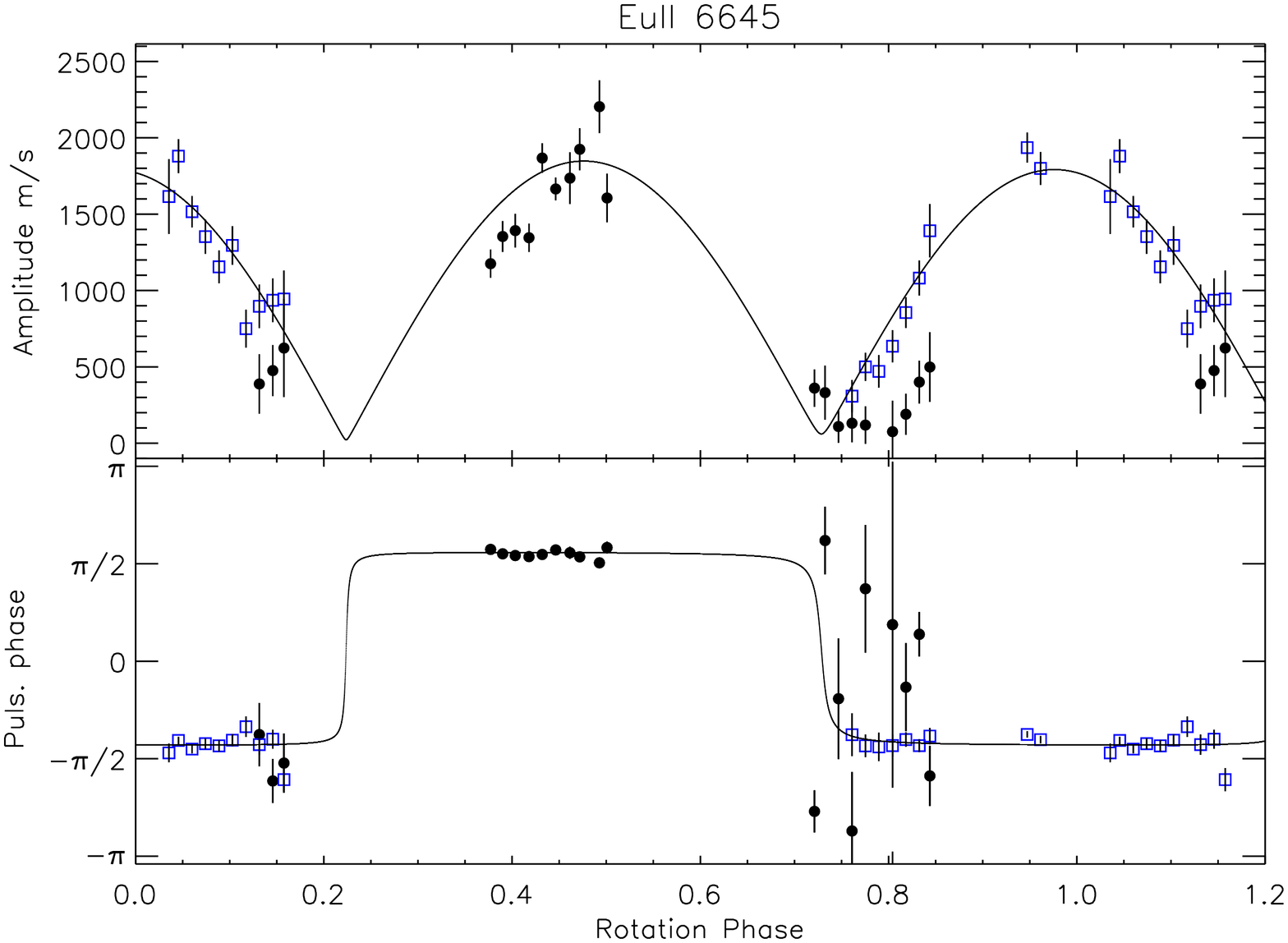}
 \caption{\label{fig:eupham} Amplitude and phase variations for features {\em A} 
(closed circles) and $B$ 
(open squares) for Eu\,\textsc{ii}\,6437.64\,\AA\ (top) and 
Eu\,\textsc{ii}\,6645.06\,\AA\ (bottom). The line profile fitting was performed 
using the CoG procedure which gives a lower amplitude than with Gaussian fitting 
(such as in Fig.\,\ref{fig:ndlpv1}), but the amplitude exceeds that 
of \halpha. The 
superposed triplet dipole model was fitted to the radial velocities of the 
strongest Eu line, Eu\,\textsc{ii}\,6645.06\,\AA. The feature $C$ (not shown) was 
also studied, but had very low amplitude and a high scatter in phases 
(consistent with its longitude) and has been 
excluded in this figure for clarity. }
\end{figure}

Angular phase-amplitude variation was studied for the Eu $A$ and $B$ features in 
the same manner as done for Nd. The phase differences between Nd and Eu of the $A$ 
and $B$ features in Figs.\,\ref{fig:ndpham} and \ref{fig:eupham} indicate 
different surface distributions such that different parts of the pulsation 
geometry are mapped, consistent with the Doppler imaging maps. 

The Eu $C$ spot was also studied for angular phase-amplitude variation, but the 
amplitude was found low and rather constant. This is not surprising, since the $C$ 
spot (see Fig.\,\ref{fig:euDI}) lies about $40^\circ$ away from the longitude of 
the pulsation pole where significantly lower pulsation amplitude is expected for a 
dipole mode. 

\begin{table*} 
\begin{minipage}{162mm} 
\caption{{\normalsize \label{tab:frq2}Pulsation amplitudes and phases ($\varphi$) 
fitted to frequencies calculated relatively to a slightly revised $\nu$ and to 
$\nu_{\rm rot}$ in Table\,\ref{tab:frq}. The data were detrended for low frequency 
trends and low frequencies. The radial velocities were measured using Gaussian 
fitting, except for lines of calcium where the CoG procedure was used. Note the 
high amplitude of the Eu lines. }} 
\begin{tabular}[ht!]{@{\,}l@{\,}r@{\,\,}r@{\,\,\,}
c@{\,\,\,}r@{\,\,}r@{\,\,\,}c@{\,\,\,}r@{\,\,}r@{\,\,\,}c@{\,\,}} \hline 
Id. 
&Amplitude~~~&$\varphi$~~~~~~~&$S/N$&Amplitude~~~&$\varphi$~~~~~~~&$S/N$&Amplitude
~~~&$\varphi$~~~~~~~&$S/N$ \\ 
 &(\ms)~~~ & (rad)~~~~ & &(\ms)~~~ & (rad)~~~~ & &(\ms)~~~ & (rad)~~~~ & \\ \hline 
 & \multicolumn{3}{c}{Ca\,\textsc{i}\,6439.07\,\AA} & 
\multicolumn{3}{c}{Eu\,\textsc{ii}\,6645.06\,\AA} & 
\multicolumn{3}{c}{Eu\,\textsc{ii}\,6437.64\,\AA} \\ 
$\nu $ &$190\pm139$&$-1.39\pm0.72$&1.4 &$ 452\pm419$ & $-1.65\pm0.91$ & 1.1 & $ 
264\pm585$ & $ 1.46\pm2.21$ &0.5 \\ 
$\nu$$-$$\nu_{\rm rot}$ &$409\pm143$&$ 1.46\pm0.35$&2.9 &$2631\pm466$ & 
$-1.59\pm0.17$ & 5.6 & $2292\pm656$ & $-1.18\pm0.28$ &3.5 \\ 
$\nu$+$\nu_{\rm rot}$ &$342\pm144$&$ 2.90\pm0.42$&2.4 &$1996\pm453$ & 
$-1.09\pm0.23$ & 4.4 & $1931\pm646$ & $-1.41\pm0.34$ &3.0 \\ 
$\nu$$-$$2\nu_{\rm rot}$ &$240\pm100$&$ 1.97\pm0.41$&2.4 &$ 377\pm320$ & $ 
0.29\pm0.85$ & 1.2 & $ 488\pm447$ & $-2.58\pm0.92$ &1.1 \\ 
$\nu$+$2\nu_{\rm rot}$ &$120\pm 99$&$ 1.36\pm0.82$&1.2 &$ 292\pm321$ & $ 
0.66\pm1.10$ & 0.9 & $ 941\pm447$ & $ 0.10\pm0.48$ &2.1 \\ 
$\nu$$-$$3\nu_{\rm rot}$ &$263\pm 91$&$ 2.62\pm0.34$&2.9 &$ 289\pm275$ & 
$-0.10\pm0.94$ & 1.0 & $ 271\pm385$ & $ 0.69\pm1.42$ &0.7 \\ 
$\nu$+$3\nu_{\rm rot}$ &$ 95\pm 91$&$-0.58\pm0.95$&1.1 &$ 215\pm274$ & $ 
2.90\pm1.27$ & 0.8 & $ 784\pm384$ & $ 2.38\pm0.49$ &2.0 \\ 
$\nu$$-$$4\nu_{\rm rot}$ &$203\pm 88$&$-2.24\pm0.44$&2.3 &$ 802\pm278$ & $ 
2.01\pm0.34$ & 2.9 & $ 120\pm383$ & $-1.99\pm3.22$ &0.3 \\ 
$\nu$+$4\nu_{\rm rot}$ &$ 11\pm 88$&$ 3.14\pm8.28$&0.1 &$ 422\pm276$ & $ 
1.25\pm0.66$ & 1.5 & $ 689\pm386$ & $-2.62\pm0.55$ &1.8 \\ 
$2\nu $ &$ 95\pm 37$&$-0.10\pm0.39$&2.6 &$ 141\pm114$ & $-2.85\pm0.81$ & 1.2 & $ 
279\pm160$ & $ 0.33\pm0.58$ &1.7 \\\vspace{.2cm} 
$2\nu$$-$$2\nu_{\rm rot}$ &$ 40\pm 37$&$-0.10\pm0.92$&1.1 &$ 299\pm114$ & $ 
1.25\pm0.38$ & 2.6 & $ 46\pm161$ & $ 2.34\pm3.50$ &0.3 \\ 
 &\multicolumn{3}{c}{Nd\,\textsc{iii}\,5102.41\,\AA} & 
\multicolumn{3}{c}{Nd\,\textsc{iii}\,6145.07\,\AA} & 
\multicolumn{3}{c}{Nd\,\textsc{iii}\,6327.24\,\AA} \\ 
$\nu $ &$281\pm 68$&$-0.29\pm0.24$&4.2 & $367\pm186$ & $-0.95\pm0.51$ & 2.0 & $ 
267\pm83 $ &$-0.99\pm0.31$& 3.2 \\ 
$\nu$$-$$\nu_{\rm rot}$ &$1000\pm 78$&$ 2.78\pm0.08$&12.7 &$1339\pm193$ & 
$-2.89\pm0.14$& 6.9 & $1439\pm86 $ &$-2.73\pm0.06$& 16.7 \\ 
$\nu$+$\nu_{\rm rot}$ &$530\pm 78$&$ 2.64\pm0.15$&6.8 & $778\pm196$ & $ 
1.72\pm0.25$ & 4.0 & $ 282\pm89 $ &$ 1.52\pm0.32$& 3.2 \\ 
$\nu$$-$$2\nu_{\rm rot}$ &$318\pm 57$&$ 1.16\pm0.18$&5.5 & $597\pm132$ & $ 
0.33\pm0.22$ & 4.5 & $ 456\pm60 $ &$ 0.42\pm0.13$& 7.6 \\ 
$\nu$+$2\nu_{\rm rot}$ &$472\pm 57$&$-2.43\pm0.12$&8.2 & $974\pm134$ & 
$-1.70\pm0.14$ & 7.3 & $ 480\pm59 $ &$-1.37\pm0.12$& 8.1 \\ 
$\nu$$-$$3\nu_{\rm rot}$ &$194\pm 47$&$-1.40\pm0.24$&4.2 & $339\pm121$ & 
$-0.89\pm0.36$ & 2.8 & $ 381\pm54 $ &$-1.44\pm0.14$& 7.1 \\ 
$\nu$+$3\nu_{\rm rot}$ &$242\pm 47$&$-1.40\pm0.19$&5.2 & $450\pm123$ & 
$-0.67\pm0.27$ & 3.7 & $ 286\pm54 $ &$-0.91\pm0.19$& 5.3 \\ 
$\nu$$-$$4\nu_{\rm rot}$ &$ 25\pm 50$&$ 0.78\pm1.97$&0.5 & $305\pm117$ & $ 
2.49\pm0.38$ & 2.6 & $ 346\pm52 $ &$ 2.84\pm0.15$& 6.6 \\ 
$\nu$+$4\nu_{\rm rot}$ &$208\pm 50$&$ 0.66\pm0.24$&4.2 & $575\pm121$ & $ 
1.75\pm0.21$ & 4.8 & $ 340\pm55 $ &$ 1.96\pm0.16$& 6.2 \\ 
$2\nu $ &$146\pm 19$&$ 1.05\pm0.13$&7.6 & $113\pm 49$ & $ 0.33\pm0.44$ & 2.3 & $ 
59\pm22 $ &$-0.04\pm0.37$& 2.7 \\ 
$2\nu$$-$$2\nu_{\rm rot}$ &$147\pm 19$&$ 0.83\pm0.13$&7.6 & $144\pm 49$ & $ 
2.77\pm0.34$ & 2.9 & $ 160\pm22 $ &$ 1.39\pm0.14$& 7.3 \\ 
\hline \hline 
\end{tabular} 
\end{minipage} 
\normalsize 
\end{table*} 
 
\section{conclusions}

We have performed a first analysis of the pulsation of HD\,99563 based on one of 
the most comprehensive spectroscopic efforts on a roAp star so far. We show that 
the pulsation is well-described by a single distorted oblique dipole mode, with 
different distortion terms for different elements. In the future, we will use 
these differences to make further inference concerning the pulsation velocity 
field -- ultimately in 3-D.  By using Doppler imaging maps of the surface 
distribution of Nd and Eu, we have identified two relatively concentrated spots 
near the magnetic poles. During rotation phases at magnetic crossover, line 
doubling in absorption features of these elements (with each component relating to 
a spot) is considerable, and we show that the stellar photosphere at opposite 
sides of the star have mutual $\pi$ rad offset, again supporting the oblique 
dipole rotator geometry. These spots allow the study of the pulsation mode for 
only parts of the visible hemisphere of the star, and from varying aspect, an 
opportunity only possible for roAp stars. Remarkably, blue-to-red running wave 
features were found only near the rotation phases where the lines were doubled and 
the spots were seen at high aspect angle. 

Our study included a short data set we also obtained with UVES on the VLT 3 years 
earlier (Elkin et al. 2005), which after a minor revision of the pulsation 
frequency showed amplitudes and phases for H, Nd and Eu \changeb{that in general 
are} consistent with 
the new data. Some roAp stars exhibit short lifetimes of pulsation modes -- in the 
most extreme case, HD\,60435, the mode lifetimes are less than a week. The 
stability of HD\,99563 over a 3-yr time span simplifies the interpretation of the 
pulsations. This also suggests that it will be possible to merge future 
observations of the rotational phases not covered by our data for a more detailed 
study. 

We found a clear rotational variation in the radial velocities of the H$\alpha$ 
line in addition to the pulsational variations which we interpret to be the first 
observational evidence of hydrogen overabundance at the magnetic/pulsational 
poles, as expected theoretically as a result of helium depletion. While we prefer 
this interpretation, alternatively, we found the first planet orbiting a roAp 
star, a hot Jupiter with a mass of 26 times the mass of Jupiter. In future work we 
will model the H$\alpha$ line radial velocity variations in detail to test our 
preferred hypothesis of hydrogen spots. 

\section*{Acknowledgments}

LMF, DWK and VGE acknowledge support for this work from the Particle Physics and 
Astronomy Research Council (PPARC) and from the Science and Technology Facilities 
Council (STFC). LMF received support from the Danish National Science Research 
Council through the project `Stellar structure and evolution -- new challenges 
from ground and space observations', carried out at Aarhus University and 
Copenhagen University. \changeb{We thank two anonymous referees for their comments that 
helped to improve the paper}. We acknowledge extensive usage of the VALD, VizieR, SIMBAD, 
ADS (NASA) databases. WZ is supported by the FP6 European Coordination Action 
HELAS and by the Research Council of the University of Leuven under grant 
GOA/2003/04.

\label{lastpage} 

\end{document}